\documentclass[letterpaper,journal]{IEEEtran}      

\usepackage[dvips]{graphicx}
\usepackage{amsmath}
\usepackage{subfigure}
\usepackage{here}
\usepackage{url}
\usepackage{cite}
\usepackage{array}
\usepackage{amssymb}
\usepackage{amsfonts}

\usepackage{ntheorem}
\makeatletter
\renewtheoremstyle{plain}
  {\item{\theorem@headerfont ##1\ ##2\theorem@separator}~}
  {\item{\theorem@headerfont ##1\ ##2\ (##3)\theorem@separator}~}
\makeatother

{\theoremheaderfont{\upshape\bfseries}
 \theorembodyfont{\normalfont\slshape}
}
\newtheorem{thm}{Theorem}
\newtheorem{algo}{Algorithm}

\newtheorem{asm}{Assumption}
\newtheorem{lem}{Lemma}
\newtheorem{prop}{Proposition}
\newtheorem{example}{Example}
\title{\LARGE \bf
Efficient PageRank Computation via Distributed Algorithms\\[1mm]
with Web Clustering}

\author{Atsushi Suzuki and Hideaki Ishii
\thanks{A.~Suzuki and H.~Ishii are with the Department of Computer Science,
Tokyo Institute of Technology, Yokohama, 226-8502, Japan.
E-mails: suzuki@sc.dis.titech.ac.jp, ishii@c.titech.ac.jp}%
\thanks{This work was supported in part by the 
JST CREST Grant No.~JPMJCR15K3 and
by JSPS under Grant-in-Aid for Scientific Research 
Grant No.~15H04020 and No.~18H01460.}}%

\begin{document}
\maketitle

\begin{abstract}
PageRank is a well-known centrality measure for the web 
used in search engines, representing the importance of 
each web page. 
In this paper, we follow the line of recent research 
on the development
of distributed algorithms for computation of PageRank, 
where each page computes its own PageRank value by
interacting with pages connected over hyperlinks.
Our approach is novel in that it is based on a 
reinterpretation of PageRank, which leads us to a set 
of algorithms with exponential convergence rates. 
We first employ gossip-type randomization 
for the page selections in the update iterations.
Then, the algorithms are generalized 
to deterministic ones, allowing simultaneous updates
by multiple pages. Finally, based on these algorithms, 
we propose a clustering-based scheme, in which groups 
of pages make updates by locally interacting among themselves
many times to expedite the convergence. 
In comparison with other existing techniques, 
significant advantages can be exhibited in their convergence 
performance, as demonstrated via numerical examples using
real web data,
and also in the limited amount of communication
required among pages.
\end{abstract}

\section{Introduction}

For search engines at Google, one of the many measures used
for ranking the web pages in search results is the so-called PageRank.
For each web page, the PageRank value provides a measure of its
importance or popularity, which is based on the network structure of
the web in terms of the hyperlinks. A page is considered more
important and popular if it receives more hyperlinks from
other pages and especially those that are important themselves.
PageRank has received a great deal of interest in the context
of complex networks as it is an effective measure of centrality; 
see, e.g., \cite{Gleich:15,IshTem:14,LanMey:06} and the references therein.

The problem of computing PageRank has been a subject of studies
over the years. Despite the simple nature of the problem, 
because of the problem size involving billions of pages in the web, 
its efficient computation remains a difficult task. 
For centralized computation, the simple power method has been
the realistic option for this reason. 
Alternative methods have been studied based on 
Monte Carlo simulations of the underlying Markov chain (e.g., \cite{AvrLitNem:07})
and distributed algorithms (e.g., \cite{SarMolPan:15,ShiYuYan:03}). 

This paper follows the line of recent research
in systems and control, where PageRank has gained much 
attention from the viewpoint of distributed algorithms.
The approach is to view each web page as an agent 
which computes its own
PageRank value iteratively by communicating with neighbors connected via hyperlinks. 
In \cite{IshTem:10,IshTem:14}, it was pointed out that the problem 
shares similarities with the multi-agent consensus 
problem \cite{Bullo:18,MesEge:10}, 
and randomized distributed algorithms were developed. To cope with the 
network size, the pages determine
to initiate updates randomly, which is called gossipping; 
for the use of randomization techniques in the systems control literature,
see \cite{TemCalDab_book}. 
The method is guaranteed to converge in the mean-square sense. 
However, it involves the time averaging of the state values,
resulting in the convergence rate
of order $1/k$ with respect to the updating time $k$. 

This approach has been further extended in different directions.
An efficient computation scheme 
based on aggregation of pages is presented 
in \cite{IshTemBai:tac12}, which provides 
another motivation of our study
as we will discuss below. 
In \cite{ZhaCheFan:tac13}, an alternative analysis 
of the algorithms was carried out based on 
methods in stochastic approximation.
Moreover, in \cite{ChaHadRab:16,IshTemBai:scl12,LeiChe:15}, 
different probability distributions are employed 
for the randomization. 
Related studies on distributed computation of PageRank include
\cite{NazPol:11,PolTre:12,RavFraTem:15}.
More in general, distributed computation of 
other network centrality measures is studied 
in, e.g., \cite{MonOliGas:18,WanTan:15}.
Other works considered the problem of optimizing PageRank 
for pages of interest by changing the link 
structure \cite{CsaJunBlo:14,FerAkiBou:tac12}
and a game theoretic analysis for enhancing PageRank via
page aggregation \cite{MaeIshAlg:17}.

More recently, distributed algorithms for PageRank demonstrating 
exponential convergence speeds
were proposed. In \cite{YouTemQiu:17}, the PageRank problem is
formulated as a least squares problem and then a gradient-based
distributed algorithm is applied. 
The algorithm in \cite{LagZacDab:17} introduces an additional
feature to maintain the state to be a probability
vector throughout the iterations. 
The work \cite{DaiFre:17} employs 
techniques from matching pursuit algorithms and
presents a randomized version. 

In this paper, we introduce a new approach for interpreting the PageRank
problem by reexamining its definition. The idea is quite simple,
but as a consequence, we arrive 
at a set of very efficient distributed 
algorithms.
We propose algorithms for both synchronous and asynchronous 
cases
in the communication among the linked pages
and fully analyze their convergence properties, which 
are shown to be exponential. 
In particular, for the asynchronous case,
we first employ randomization-based gossipping, but then
extend the approach to deterministic gossipping, where
multiple pages may be selected to simultaneously 
make updates. As long as each page updates its state
infinitely often, 
convergence to its corresponding PageRank value
is guaranteed.

The highlight of this work is that through this development, 
we become able to construct an efficient algorithm based on 
clustering of web pages for distributed computation of Page\-Rank.
As the web inherently has a hierarchical structure, clustering 
can be easily carried out,
for example, by grouping pages in the same domains or subdomains. 
In this context, instead of the pages, it is the groups 
that initiate updates for their member pages. 
When a group determines to do so,
the pages in the group make calculations
by interacting among themselves, 
which is equivalent to iterating infinitely many times. 
Such updates can actually be performed in one step, expressed as
matrix operations
involving only local states. Part of the computation
can be completed offline based on the information
of the link structure within the group. Hence, 
the additional requirement for
computation should be limited. 
We demonstrate the fast convergence performance
in numerical examples using real web data. 

The novel aspects of our approach can be summarized
as follows. First, the reformulation idea is simple
and its advantage may not be immediately clear.
This is partly because additional states are introduced
for the pages, which may increase the computational burden.
In fact, in the synchronous case, the convergence is 
not necessarily faster than the power method. 
Second, in the proposed algorithms, the states 
are guaranteed to reach the true PageRank values from 
below in a monotonic fashion. Hence, even if randomization
is adopted, the responses of the states are smooth, which 
may explain the efficiency of the approach. 
Third, the pages communicate only over their
outgoing hyperlinks and do not require the knowledge of
the incoming ones; this is another advantage of the
schemes in comparison with conventional methods.






In the clustering-based algorithm, pages within each
group collectively update their values and the exact values
of Page\-Rank can be obtained. 
It is emphasized that 
this approach
relies on the properties of the specific schemes 
developed in this paper. 
In gossipping for page selections, no specific 
randomization is required especially for 
obtaining the true PageRank.
Moreover, multiple pages can make updates at the same time
partly due to the simple communication scheme. 

It is remarked that for large-scale computation, a related, but slightly 
different approach based on web aggregation has been 
studied as well.
There, aggregated PageRank values representing the groups
are computed first, and then a more local computation 
takes place within groups
to assign values to individual pages, which typically results
in approximation in the final values. 
Such studies can be found, for example, 
in \cite{LanMey:06,ZhuYeLi:05} using classical 
methods in Markov chains and in \cite{BroLem_infret:06} 
via extensive simulations. 
The work \cite{IshTemBai:tac12} developed a method motivated 
by the studies on large-scale systems based on singular perturbation 
analyses for Markov chains (e.g., \cite{AldKha:91}) and
consensus networks (e.g., \cite{BiyArc:08}).

This paper is organized as follows: 
In Section~\ref{sec:pagerank}, 
we first give a brief overview of the PageRank problem 
and then introduce an alternative formulation.
In Section~\ref{sec:dist}, a novel distributed
algorithm based on randomized gossipping 
is presented along with an analysis
on their convergence properties. 
In Section~\ref{sec:generalized}, we extend our approach
and develop a generalized distributed algorithm. 
This is then further
exploited to deal with clustering-based calculations in
Section~\ref{sec:group}.
Illustrative numerical examples are provided in Section~\ref{sec:example}.
The paper is finally concluded in Section~\ref{sec:concl}.
Preliminary versions of this paper have appeared 
as \cite{SuzIsh:acc18,SuzIsh:cdc19}. The current paper
provides the full proofs of the results and extended
discussions along with a numerical example of larger scale. 

\smallskip
\noindent
{\it Notation}:~~For vectors and matrices, inequalities 
are used to denote entry-wise inequalities:
For $X,Y\in\mathbb{R}^{n\times m}$, $X\leq Y$ implies
$x_{ij}\leq y_{ij}$ for all $i,j$; 
in particular, we say that the matrix $X$ is nonnegative if $X\geq 0$ 
and positive if $X> 0$.
A probability vector is a nonnegative vector $v\in\mathbb{R}^n$ such that
$\sum_{i=1}^n v_i = 1$.
A matrix $X\in\mathbb{R}^{n\times n}$ is said to be (column) stochastic
if it is nonnegative and each column sum equals 1, i.e.,
$\sum_{i=1}^n x_{ij}=1$ for each $j$.
Let $\mathbf{1}_n\in\mathbb{R}^n$ be the vector whose entries are all $1$ as
$\mathbf{1}_n:=[1\,\cdots\,1]^T$.
For a discrete set $\mathcal{D}$, its cardinality
is given by $\bigl|\mathcal{D}\bigr|$.

\section{A Novel Approach Towards PageRank}\label{sec:pagerank}

In this section, we briefly introduce the notion of PageRank
and its interpretation commonly employed for its computation.
Then, we discuss an alternative formulation of the problem,
which will lead us to a novel class of distributed algorithms. 

\subsection{The PageRank Problem}

The computation of PageRank proposed by Brin and Page 
\cite{BriPag:98} 
starts with regarding the entire web as a directed graph. 
Let $n$ be the number of pages; we assume $n\geq 2$ to avoid the trivial case.
The web graph is given by $\mathcal{G}:=\left( \mathcal{V}, \mathcal{E}\right)$ 
where $\mathcal{V}:=\left\{ 1,2,\ldots,n \right\}$ is the 
set of vertices representing the web pages, and $\mathcal{E}$ is 
the set of hyperlinks connecting the pages. 
Here, $(i,j)\in \mathcal{E}$ holds if and only if page $i$ has a hyperlink to page $j$.
Hyperlinks are not always mutual, so this graph is generally a directed graph.
For node~$i$, let 
the set of outgoing neighbors and
that of incoming neighbors be given, respectively, by
$\mathcal{L}_{i}^{\text{out}}:=\left\{ j:\,(i,j)\in\mathcal{E} \right\}$ and
$\mathcal{L}_{i}^{\text{in}}:=\left\{ j:\,(j,i)\in\mathcal{E} \right\}$. 

When a node does not have any outgoing link, it is referred to as a dangling node.
Here, to simplify the discussion, we assume that all pages have at least 
one outgoing hyperlink. 
This is commonly done by slightly modifying the structure of the web,
specifically by adding hyperlinks from such dangling nodes, which 
correspond to the use of back buttons; see, e.g., \cite{LanMey:06,LeiChe:15} for more details.

Next, we define the hyperlink matrix $A=\left( a_{ij} \right)\in \mathbb{R}^{n\times n}$ 
of this graph by
\begin{equation*}
  a_{ij} 
   := \begin{cases}
	\frac{1}{n_j} & \text{if $i\in\mathcal{L}_j^{\text{out}}$},\\
	0 & \text{otherwise},
      \end{cases}
\end{equation*}
where $\mathcal{L}_{i}^{\text{out}}$ is the set 
of outgoing neighbors of page~$i$ and 
$n_i$ is its cardinality. 
By the assumption that all pages have one or more hyperlinks, 
this matrix $A$ is stochastic.


For the web consisting of $n$ pages, the PageRank vector $x^*\in \mathbb{R}^n$ 
is defined as
\begin{equation}
  x^{*} 
   = (1-m)Ax^{*} + \frac{m}{n}\mathbf{1}_n,
      ~~\mathbf{1}_n^{T}x^{*}=1,
\label{def-normal}
\end{equation}
where the parameter is chosen as $m\in(0,1)$;
in this paper, we take the commonly used value $m=0.15$.

The definition in \eqref{def-normal} can be rewritten as 
	\begin{equation}
		x^{*}=M x^{*},~~\mathbf{1}_n^{T}x^{*}=1,
		\label{def-normal-mod}
	\end{equation}
where the modified link matrix $M$ is given by 
$M=(1-m)A + (m/n) \mathbf{1}_n\mathbf{1}_n^{T}$.
Since $M$ is a convex combination of two stochastic matrices
$A$ and $(1/n)\mathbf{1}_n\mathbf{1}_n^{T}$,
it is stochastic as well. It is now clear that $x^*$ is the
eigenvector corresponding the eigenvalue 1 of the link matrix $M$. 

For its computation, the PageRank vector $x^*$ can be obtained by solving 
the linear equation \eqref{def-normal} or \eqref{def-normal-mod}. 
However, due to its large dimension, the computation must rely on simple
algorithms. It is common to use the power method given by the iteration
of the form
\begin{equation}
  x(k+1) = (1-m)Ax(k) + \frac{m}{n}\mathbf{1}_n,
\label{power}
\end{equation}
where $x(k)\in\mathbb{R}^n$ is the state whose initial value $x(0)$
can be taken as any probability vector. By Perron's theorem \cite{HorJoh:85},
it follows that $x(k)\to x^*$ as $k\to \infty$.

\begin{figure}[tb]
        \vspace*{1mm}
	\centering
	\includegraphics[width=6cm]{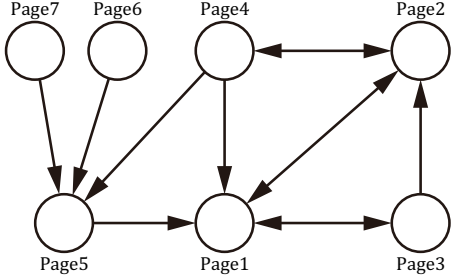}
        \vspace*{-7mm}
	\caption{An example graph with seven nodes}
	\label{example1}
\end{figure}

\begin{example}\label{example2}\rm
Consider the web consisting of seven pages depicted in 
Fig.~\ref{example1}.
%
%
%
%
	We can calculate the PageRank vector of this graph as
	\begin{align*}
		x^{*}&=\bigl[
			0.316 ~
			0.259 ~
			0.156 ~
			0.132 ~
			0.0951 ~
			0.0214 ~
			0.0214
		\bigl]^T.
	\end{align*}
Note that the indices of the pages are given according 
to the order of their PageRank values. 
Pages~1 and~2 rank the first and second, which can be
due to having, respectively, 4 and~3 incoming links. 
Pages~3 and~4 have only 1 incoming link, but
take better rankings than page~5, which has 3~links. 
This is because the ranks depend not only on the number
of incoming links, but also on the values of the pages 
from which the links originate. 
In this respect, pages~3 and~4 are clearly advantageous,
being neighbors of pages~1 and~2.
Pages~6 and~7 have no incoming hyperlink and, as a result, 
take the lowest possible value, which is equal to $m/n=0.15/7= 0.0214$. 
%
\end{example}

\subsection{Reformulation of the PageRank Problem}\label{kasaneawase}

Now, we present a new formulation of Page\-Rank by transforming its 
original definition. This formulation becomes the key for 
developing novel distributed algorithms. 
The idea itself is simple, but its advantage 
in the context of distributed computation of 
PageRank will become clear. 

The formula of PageRank in (\ref{def-normal}) can be transformed as 
\begin{align}
 & x^{*}
    = (1-m)Ax^{*} + \frac{m}{n}\mathbf{1}_n 
   \nonumber\\
 & \Longleftrightarrow ~  
    x^{*}
     = \left( I-(1-m)A \right)^{-1}\frac{m}{n}\mathbf{1}_n	
    \nonumber\\
 & \Longleftrightarrow ~ 
    x^{*}
      =	\displaystyle 
         \sum_{t=0}^{\infty}\left( (1-m)A \right)^{t} 
             \frac{m}{n}\mathbf{1}_n.
\label{eq_kasaneawase}
\end{align}
In the last transformation, the Neumann series 
(e.g., \cite{HorJoh:85}) is applied. 
Notice that $(1-m)A$ is a Schur stable matrix because the link matrix $A$ is
stochastic with spectral radius equal to 1.

The formula in \eqref{eq_kasaneawase} implies that the PageRank
computation can be carried out iteratively in several ways. 
It is immediate to write down an equation for the state $x(k)\in\mathbb{R}^n$ 
given by
\begin{equation}
  x(k) = \displaystyle \sum_{t=0}^{k}\left( (1-m)A \right)^{t} 
            \frac{m}{n}\mathbf{1}_n.
\label{eqn:alt_formula}
\end{equation}
Clearly, the power method in \eqref{power} is a compact way to realize
this using only $x(k)$ as the state. 

Another approach is to use a slightly redundant iteration by using
an additional state. This is denoted by $z(k)\in\mathbb{R}^n$. 
Set the initial states as 
$x(0)=z(0)=(m/n)\mathbf{1}_n$.
Then, the update scheme of the two states is given as follows:
\begin{equation}
\begin{split}
  x(k+1) &= x(k) + (1-m)Az(k),\\
  z(k+1) &= (1-m)Az(k).
\end{split}
\label{eqn:synch}
\end{equation}
Through this alternative algorithm, we can obtain the 
Page\-Rank vector $x^*$.
We formally state this along with its
other properties as 
a lemma in the following. Similar properties will appear in our development of distributed algorithms later.

\begin{lem}\rm\label{lemm:synch}
In the update scheme in \eqref{eqn:synch}, the states $x(k)$ and
$z(k)$ satisfy the following:
\begin{enumerate}
\item[(i)] $z(k)\rightarrow 0$ as $k\rightarrow\infty$.
\item[(ii)] $x(k)\leq x(k+1)\leq x^*$ for $k$.
\item[(iii)] $x(k)\rightarrow x^*$ as $k\rightarrow\infty$.
\end{enumerate}
\end{lem}

\textit{Proof:}
(i)~As the link matrix $A$ is stochastic, its spectral radius 
equals 1, and thus $(1-m)A$ is a Schur stable matrix. 
This implies that $z(k)$ converges to zero. 

(ii)~Note that $z(k)\geq 0$ because $A$ is stochastic and $z(0)>0$. 
Furthermore, we have $x(0)>0$. Thus, it is clear that $x(k)$ is 
nondecreasing as a function of $k$. The fact that it is upper bounded by $x^*$
follows from (iii), which is shown next.

(iii)~From \eqref{eqn:synch}, we can write $x(k)$ as
\begin{align}
 x(k)	
  &= x(0) + \displaystyle \sum_{t=1}^{k} z(t)
   = x(0) + \displaystyle \sum_{t=1}^{k}
            \left( (1-m)A \right)^{t} z(0)
  \nonumber\\
  &= \displaystyle \sum_{t=0}^{k}
            \left( (1-m)A \right)^{t} \frac{m}{n}\mathbf{1}_n.
 \label{eqn:prop:synch}
\end{align}
This and \eqref{eq_kasaneawase} indicate that 
$x(k)$ converges to $x^{*}$ as $k\to \infty$.
\hfill\mbox{$\blacksquare$}

We have a few remarks on the alternative approach introduced above
in comparison with the power method in \eqref{power}. 
First, the computation requires the second state $z(k)$ in addition to $x(k)$. 
As seen in \eqref{eqn:prop:synch}, this state $z(k)$ is integrated over time
to compute $x(k)$ in \eqref{eqn:alt_formula}.
Second, the initial values of $x(k)$ and $z(k)$ are fixed to 
$(m/n)\mathbf{1}_n$,
and there is no freedom in these choices. Hence, each time the computation
takes place through the update scheme \eqref{eqn:synch}, the algorithm 
cannot make use of the PageRank values computed in the past as initial guesses.
This point may be a limitation of this approach. 
Also, the initial states are not probability vectors
as in the power method. In fact, $x(k)$ becomes a probability vector only after
converging to $x^*$.
Third, notice that $m/n$ is the minimum PageRank value, which will be
assigned to pages having no incoming links. For such pages,
the states will not change during the updates. 

\subsection{Distributed Algorithm for Synchronous Updates}

One interpretation of \eqref{eqn:synch} from the perspective of distributed algorithms
can be given as follows:
\begin{enumerate}
\item At time $0$, all pages start with the value ${m}/{n}$.
\item At time $k$, each page attenuates its current value by $1-m$ and then sends 
it to its linked pages after equally dividing it.
At that time, page $i$ computes the weighted sum of the values received from
the neighbors having links to the page.
\end{enumerate}

Though we do not discuss in this paper, there is a generalized PageRank 
definition which uses a probability vector $v\in\mathbb{R}^n$ 
instead of $(1/{n})\mathbf{1}_n$, 
that is, $x^*=(1-m)Ax^*+mv$. In such a case, the proposed algorithm can be
modified by replacing the initial states with $x(0)=z(0)=mv$.


We finally present a distributed algorithm based on \eqref{eqn:synch}.



\begin{algo}[Synchronous Distributed Algorithm]\label{alg:1}\rm
\begin{enumerate}
\item For page $i$, set the initial values as $x_{i}(0)=z_{i}(0)=m/n$.
\item At time $k$, page $i$ transmits its value $z_i(k)$ to its neighbors along
its outgoing links and then
updates its states to obtain $x_i(k+1)$ and $z_i(k+1)$ according to
\begin{align*}
  x_{i}(k+1)
   &= x_i(k) + \displaystyle \sum_{j:\,i\in \mathcal{L}_j^{\text{out}}}\frac{1-m}{n_j}z_{j}(k),\\
  z_{i}(k+1) 
   &= \displaystyle \sum_{j:\,i\in \mathcal{L}_j^{\text{out}}}\frac{1-m}{n_j}z_{j}(k).
\end{align*}
\end{enumerate}
\end{algo}

As we show through simulations in Section~\ref{sec:example}, 
this synchronized algorithm may not be particularly fast, especially
in comparison with the power method. Moreover, due to the additional
state $z(k)$, the algorithm requires more memory. 
The advantage of the proposed
reformulation however becomes evident in the asynchronous versions
of this distributed algorithm, which will be presented in 
the next section.

\section{Gossip-Type Distributed Algorithms}\label{sec:dist}

We now develop asynchronous versions of the distributed algorithm.
They are based on randomized communication among the pages, 
which is referred to as gossipping. 

In the asynchronous update schemes, at each time $k$, 
one page is randomly chosen, which transmits its current state value
to the linked pages. We present two algorithms which differ in
their probability distributions for selecting the updating pages. 
One uses the uniform distribution and the other is more general.
In both cases, the distributions remain fixed throughout the 
execution of the algorithms; thus, the updating
pages are chosen in an independently identically distributed (i.i.d.) manner.
Denote by $\theta(k)\in\mathcal{V}$ the selected page at time $k$. 
In the section on numerical examples, 
it will be shown that nonuniform distribution may 
be beneficial from the perspective of convergence speed.

\subsection{Algorithm Based on the Uniform Distribution}\label{unif-algo}

We consider the case where the selection of the updating 
pages follows the uniform distribution. 
The proposed distributed algorithm for this case is provided below.

\begin{algo}[Distributed Randomized Algorithm]\label{alg:2}\rm
\begin{enumerate}
\item For page $i\in\mathcal{V}$,
set the initial values as $x_{i}(0)=z_{i}(0)={m}/{n}$.
\item At time $k$, 
select one page $\theta(k)$ based on the uniform distribution: 
\begin{equation}
  \text{Prob}\bigl\{ 
               \theta(k)=i  
             \bigr\} = \frac{1}{n}~~\text{for $i\in\mathcal{V}$}. 
\label{uniform}
\end{equation}
  \item Page $\theta(k)$ transmits its value $z_{\theta(k)}(k)$ 
over its outgoing links to pages in $\mathcal{L}^{\text{out}}_{\theta(k)}$.
  \item Each page $i\in\mathcal{V}$ updates its states
        to obtain $x_i(k+1)$ and $z_i(k+1)$ as
	\begin{align}
	x_{i}(k+1)
 	 &= \begin{cases}
		x_i(k)+\frac{1-m}{n_{\theta(k)}}z_{\theta(k)}(k)
                & \text{if $i\in\mathcal{L}^{\text{out}}_{\theta(k)}$},\\
		x_i(k) & \text{otherwise,}
            \end{cases}\notag\\
	 z_{i}(k+1)
           &= \begin{cases}
		0 & \text{if $i=\theta (k)$},\\
		z_i(k) + \frac{1-m}{n_{\theta(k)}} z_{\theta(k)}(k)
                    & \text{if $i\in\mathcal{L}^{\text{out}}_{\theta(k)}$},\\
		z_i(k) & \text{otherwise.}
              \end{cases}
	\label{eqn:gdra1}
	\end{align}
\end{enumerate}
\end{algo}

The resemblance of this algorithm to Algorithm~\ref{alg:1} is obvious.
The two states $x_i(k)$ and $z_i(k)$ play similar roles in both
algorithms. 
The differences are that in the asynchronous case, 
the updates are made with one neighbor at a time, and 
also both $x_i(k)$ and $z_i(k)$ are integrated
over time. For $z_i(k)$, this was not the case in Algorithm~\ref{alg:1}. 
The two variables are updated differently when page~$i$ 
is the selected page $\theta(k)$ at time $k$ as in such cases, 
its own $z_i(k)$ is set to zero. By contrast, in Algorithm~\ref{alg:1}, 
$z_i(k)$ is zero only in the case where page $i$ has no incoming link.

We now rewrite this algorithm in a vector form.
First, let 
\begin{equation}
   Q := (1-m)A. 
\label{eqn:Q}
\end{equation}
Denote by $e_i$ and $q_i$, respectively, 
the $i$th columns of 
the $(n \times n)$-identity matrix and $Q$
for $i\in\mathcal{V}$.
Then, we define the matrices $Q_i,R_i,S_i\in\mathbb{R}^{n\times n}$,
$i\in\mathcal{V}$, by
\begin{equation}
\begin{split}
 Q_i &:= \begin{bmatrix}
            e_1 & e_2 & \cdots & e_{i-1} & q_i & e_{i+1} &\cdots& e_{n}  
         \end{bmatrix},\\
 R_i &:= \begin{bmatrix}
            {0}_n & {0}_n & \cdots& {0}_n & q_i 
             & {0}_n & \cdots & {0}_n 
         \end{bmatrix},\\
  S_i 
   &:= \begin{bmatrix}
         e_1 & e_2 & \cdots & e_{i-1} & {0}_n 
          & e_{i+1} & \cdots & e_{n}  
      \end{bmatrix},
\end{split}
\label{eqn:QiRi}
\end{equation}
where in both $Q_i$ and $R_i$, it is the $i$th column 
that is equal to $q_i$,
and in $S_i$, only the $i$th column is zero. 
Note that the matrices $Q$, $Q_i$, $R_i$, and $S_i$
are all nonnegative matrices for $i\in\mathcal{V}$.
Moreover, by definition, it holds $Q_i = R_i+S_i$.  

As in the synchronous case, 
the initial states are taken as $x(0)=z(0)=({m}/{n})\mathbf{1}_n$. 
The update schemes in \eqref{eqn:gdra1} for the two states can be 
written in the compact form as
\begin{align}
\begin{split}
  x(k+1) &= x(k) + R_{\theta(k)} z(k),\\
  z(k+1) &= Q_{\theta(k)} z(k),
\end{split}
\label{eqn:gdra1v}
\end{align}
where $\theta(k)$ is the page selected for updating
at time $k$ in step~2) of the algorithm.

We are now ready to present the main result for this distributed algorithm for PageRank computation. It shows that the true PageRank values can be obtained almost surely.

\begin{thm}\label{unif_thm}\rm
  Under Algorithm~\ref{alg:2}, the PageRank vector $x^*$ is computed 
  with $x(k)\to x^{*}$ as $k\to \infty$ with probability $1$. 
%
%
In particular, the following two properties hold:
\begin{enumerate}
 \item[(i)] $x(k)\leq x(k+1)\leq x^{*}$ holds for $k\geq 0$.
 \item[(ii)] $\mathbb{E}\left[ x(k)\right]\to x^{*}$ as $k\to \infty$, and the
 convergence speed is exponential. 
\end{enumerate}
\end{thm}

We first show the following lemma
regarding the synchronous update scheme 
\eqref{eqn:synch}. 
It is a simple result, but will be useful in the proofs
of different results in this paper. 
Denote by $x_{\rm{s}}(k)$ and $z_{\rm{s}}(k)$ 
the states in \eqref{eqn:synch}. 
Let 
\begin{equation}
  \widetilde{Q}_{\rm{s}}(k) 
    := Q^{k+1}.
\label{eqn:Qtilde}
\end{equation}
Also, let $q_{ij}:=\left[ Q \right]_{ij}=(1-m)a_{ij}$ 
for $i,j\in\mathcal{V}$.
Clearly, $q_{ij}\geq 0$ holds for any $i,j$.

\begin{lem}\label{lem:synch}\rm
The state $x_{\rm{s}}(k)$ of the synchronous update scheme 
\eqref{eqn:synch} can be expressed as
\begin{equation}
 x_{\rm{s}}(k)
  = \frac{m}{n} \mathbf{1}_n
     + \displaystyle \sum_{t=0}^{k-1} 
      \widetilde{Q}_{\rm{s}}(t) \frac{m}{n} \mathbf{1}_n,
    ~~k\geq 0.
 \label{eqn:x_s}
\end{equation}
Moreover, the $(i,j)$th element of 
the matrix $\widetilde{Q}_{\rm{s}}(t)$ in \eqref{eqn:Qtilde}
can be written as
\begin{equation}
   \bigl[ \widetilde{Q}_{\rm{s}}(t) \bigr]_{ij}
     = 
        \sum_{m_1,\ldots,m_{t}\in\mathcal{V}}
              q_{i m_t}q_{m_t m_{t-1}}\cdots q_{m_{1}j}
  ~~\text{for $i,j$}.
\label{eqn:Qs_ij}
\end{equation}
\end{lem}

In the summation in \eqref{eqn:Qs_ij}, it is clear that 
the term $q_{i m_t}q_{m_t m_{t-1}}\cdots q_{m_{1}j}$ 
is nonzero if and only if there exists a sequence of hyperlinks
$(j,m_1)$, $(m_1,m_2),\ldots$, $(m_t,m_{t-1})$, $(i,m_t)\in\mathcal{E}$
in the web graph. It is however noted that 
such a property will not be explicitly used 
in our analysis.

\textit{Proof:} 
From the update scheme in \eqref{eqn:synch} and
the definition of $Q$ in \eqref{eqn:Q}, we have
\begin{align*}
  x_{\rm{s}}(k+1) &= x_{\rm{s}}(k) + Q z_{\rm{s}}(k),~~
  z_{\rm{s}}(k+1) = Q z_{\rm{s}}(k).
\end{align*}
Thus, it follows that 
$z_{\rm{s}}(k)
  = Q^k z_{\rm{s}}(0)$.
Furthermore, 
\[
 x_{\rm{s}}(k) 
  = x_{\rm{s}}(0) 
     + \sum_{t=0}^{k-1} Q z_{\rm{s}}(t)
  = x_{\rm{s}}(0) 
     + \sum_{t=0}^{k-1} Q^{t+1} z_{\rm{s}}(0).
\]
Since the initial states are set as
$x_{\rm{s}}(0)= z_{\rm{s}}(0)=(m/n)\mathbf{1}_n$,
we obtain \eqref{eqn:x_s}.

Because $\widetilde{Q}_{\rm{s}}(t)=Q^{t+1}$, we have that
$\bigl[ \widetilde{Q}_{\rm{s}}(t) \bigr]_{ij}$ is the summation of all terms that can 
be expressed as 
the product $q_{i m_t}q_{m_t m_{t-1}}\cdots q_{m_{1}j}$ of 
$t+1$ elements
in the matrix $Q$, where $m_1,m_2,\ldots,m_{t}$ are all 
taken from $\mathcal{V}$. Thus, we have \eqref{eqn:Qs_ij}.
\hfill\mbox{$\blacksquare$}

\smallskip
\textit{Proof of Theorem~\ref{unif_thm}:}
(i)~Since both $Q_{\theta(k)}$ and $R_{\theta(k)}$ are nonnegative, 
by \eqref{eqn:gdra1v}, $x(k)$ and $z(k)$ are nonnegative at all $k$.
In particular, $x(k)$ is a nondecreasing function of time, i.e.,
$x(k)\leq x(k+1)$ holds.

Next, we show $x(k)\leq x^*$.
This is done by proving 
\begin{equation}
 x(k)\leq x_{\rm{s}}(k)~~\text{for $k\geq 0$},
 \label{eqn:unif_thm:2a}
\end{equation}
where $x_{\rm{s}}(k)$ is the state of the synchronous 
update scheme in \eqref{eqn:synch}. 
Then, by Lemma~\ref{lemm:synch}\,(ii), 
we obtain $x(k)\leq x_{\rm{s}}(k)\leq x^{*}$.

For $k=0$, we have $x(0)= x_{\rm{s}}(0)$ and thus 
\eqref{eqn:unif_thm:2a} holds.
For $k\geq 1$, the state $z(k)$ in \eqref{eqn:gdra1v} can be written as
\begin{equation}
  z(k) 
   = Q_{\theta (k-1)}Q_{\theta (k-2)}\cdots Q_{\theta (0)}z(0). \nonumber
\end{equation}
Thus, we can express $x(k)$ as
\begin{align}
  &x(k)
   = x(0) + \displaystyle \sum_{t=0}^{k-1} R_{\theta (t)}z(t)\nonumber\\
   &~~= \displaystyle\frac{m}{n}\mathbf{1}_n 
         + \displaystyle \sum_{t=0}^{k-1}
              R_{\theta (t)}Q_{\theta (t-1)}Q_{\theta (t-2)}\cdots Q_{\theta (0)}
              \frac{m}{n}\mathbf{1}_n.
 \label{eqn:xk}
\end{align}
For $k\geq 0$, define the nonnegative matrix
$\widetilde{Q}_{\theta}(k)\in\mathbb{R}^{n\times n}$ by 
\begin{equation}
  \widetilde{Q}_{\theta}(k) 
    := R_{\theta (k)}Q_{\theta (k-1)}Q_{\theta (k-2)}
               \cdots Q_{\theta (0)}.
\label{eqn:Qthtea}
\end{equation}
Then, we have from \eqref{eqn:xk}
\begin{align}
 x(k)
  &= \frac{m}{n} \mathbf{1}_n
      + \displaystyle \sum_{t=0}^{k-1} \widetilde{Q}_{\theta}(t) \frac{m}{n} 
         \mathbf{1}_n.
 \label{eqn:x_unif_q}
\end{align}

By comparing \eqref{eqn:x_unif_q} with 
\eqref{eqn:x_s} in Lemma~\ref{lem:synch},
we observe that for establishing \eqref{eqn:unif_thm:2a},
it suffices to show the inequality below:
\begin{equation}
  \displaystyle \sum_{t=0}^{k-1} \widetilde{Q}_{\theta}(t) 
   \leq \displaystyle \sum_{t=0}^{k-1} \widetilde{Q}_{\rm{s}}(t)~~\text{for $k\geq 1$}.
\label{eqn:QQ_ineq}
\end{equation}
In particular, we should show 
\begin{equation}
 \bigl[
    \widetilde{Q}_{\theta}(t)
 \bigr]_{ij}
   \leq 
      \bigl[
         \widetilde{Q}_{\rm{s}}(t)
      \bigr]_{ij}
       ~\text{for $i,j$ and $t=0,\ldots,k-1$}.
\label{qneq}
\end{equation}
The approach for its proof is to establish that any 
term appearing on the left-hand side in \eqref{qneq}
always appears in the right-hand side. 
Since all terms are nonnegative, 
the inequality \eqref{qneq} implies \eqref{eqn:QQ_ineq}.
%
The right-hand side of \eqref{qneq}
is written out in \eqref{eqn:Qs_ij}
of Lemma~\ref{lem:synch}. 

In what follows,
we obtain the formula for the left-hand side of
\eqref{qneq}, that is,
$\bigl[ \widetilde{Q}_{\theta}(t) \bigr]_{ij}$.
Recall that by \eqref{eqn:QiRi}, we have $Q_i = R_i+S_i$.  
Hence, by using $R_i$ and $S_i$, we can write $\widetilde{Q}_{\theta}(t)$ as 
\begin{align*}
  \widetilde{Q}_{\theta}(t) 
   &= R_{\theta (t)}Q_{\theta (t-1)} Q_{\theta (t-2)}
               \cdots Q_{\theta (0)}\notag\\
   &= R_{\theta (t)}\left( R_{\theta (t-1)} + S_{\theta(t-1)} \right)
         \cdots\left( R_{\theta (0)} + S_{\theta(0)} \right).
\end{align*}
We must compute the products of $R_i$ and $S_i$ appearing on 
the far right-hand side above. 

To this end, we derive a formula for the product
where $R_i$ appears $\ell\leq t+1$ times and 
$S_i$ appears $k_{\ell}-{\ell}\leq t$ times:
\begin{equation}
 R_{\theta (k_{\ell})}\underbrace{\cdots}_{\text{$0$ or more $S_i$}}
  R_{\theta (k_{{\ell}-1})}\underbrace{\cdots}_{\text{$0$ or more $S_i$}}\cdots  
  R_{\theta(k_1)}\underbrace{\cdots}_{\text{$0$ or more $S_i$}},
 \label{eqn:prod1}
\end{equation}
where $0\leq k_1< \cdots< k_{\ell}=t\leq k-1$.
By \eqref{eqn:QiRi}, in $R_i$, all elements except 
the $i$th column are $0$ while $S_i$ is equal to
the identity matrix except for the $i$th column, which 
is a zero vector.
These facts lead us to the following relation 
for arbitrary $i,j$:
\begin{align*}
	R_{i} S_{j}&=
	\begin{cases}
		0 & \text{if $i=j$},\\
		R_{i} & \text{otherwise},
	\end{cases}
\end{align*}
Thus, the product in \eqref{eqn:prod1} becomes zero if
one of the following conditions holds:
\begin{itemize}
 \item $\theta(k_{\ell})$ is equal to one of
       $\theta(k_{\ell}-1),\ldots,\theta(k_{{\ell}-1}+1)$;
 \item $\theta(k_{{\ell}-1})$ is equal to one of
	$\theta(k_{{\ell}-1}-1),\ldots,\theta(k_{{\ell}-2}+1)$;
 \item $\cdots$
 \item $\theta(k_1)$ is equal to one of
	$\theta(k_1-1),\ldots,\theta(0)$.
\end{itemize}
Otherwise, the term in \eqref{eqn:prod1} reduces to 
\begin{align}
 &R_{\theta (k_{\ell})}\underbrace{\cdots}_{\text{$0$ or more $S_i$}}
  R_{\theta (k_{{\ell}-1})}\underbrace{\cdots}_{\text{$0$ or more $S_i$}}\cdots  
  R_{\theta(k_1)}\underbrace{\cdots}_{\text{$0$ or more $S_i$}}\notag\\
 &~~= R_{\theta (k_{\ell})} R_{\theta (k_{{\ell}-1})} \cdots R_{\theta(k_1)}.
 \label{eqn:prod2}
\end{align}
For the product of $R_i$ above, we can use the formula
\begin{align*}
  R_{i} R_{j}
   &= q_{ij} R_{j}^{(i)},
\end{align*}
where the matrix $R_{j}^{(i)}\in\mathbb{R}^{n\times n}$ 
is nonzero only in the $j$th column as
\begin{equation*}
  R_{j}^{(i)}
    := \begin{bmatrix}
            {0}_n & {0}_n & \cdots& {0}_n & q_i
             & {0}_n & \cdots & {0}_n 
         \end{bmatrix}.
\end{equation*}
We also need another formula that holds for arbitrary $i,j,m$:
\[
  R_{j}^{(i)} R_m = q_{jm} R_{m}^{(i)}.
\]
Then, applying these formulae to the product in \eqref{eqn:prod2}
repeatedly yields
\begin{align*}
 &R_{\theta (k_{\ell})} R_{\theta (k_{{\ell}-1})} \cdots R_{\theta(k_1)} \notag\\
 &~= q_{\theta(k_{\ell})\theta(k_{{\ell}-1})}q_{\theta(k_{\ell}-1)\theta(k_{{\ell}-2})}\cdots 
                    q_{\theta(k_2)\theta(k_1)} R_{\theta(k_1)}^{(\theta(k_{\ell}))}. 
\end{align*}
Finally, 
the $(i,j)$th element is obtained as
\begin{align*}
 &\left[
     R_{\theta (k_{\ell})} R_{\theta (k_{{\ell}-1})} \cdots R_{\theta(k_1)}
  \right]_{ij} \notag\\
 &~= q_{\theta(k_{\ell})\theta(k_{{\ell}-1})}
     q_{\theta(k_{\ell}-1)\theta(k_{{\ell}-2})}\cdots 
                    q_{\theta(k_2)\theta(k_1)}
     \left[
         R_{\theta(k_1)}^{(\theta(k_{\ell}))}
     \right]_{ij} \notag\\
 &~= \begin{cases}
         q_{i\,\theta(k_{\ell})}
         q_{\theta(k_{\ell})\theta(k_{{\ell}-1})}
         q_{\theta(k_{\ell}-1)\theta(k_{{\ell}-2})}\cdots 
                    q_{\theta(k_2)\,j}\\
         ~~~~~~\text{if $\theta(k_1)=j$},\\
         0~~~~\text{otherwise}.
      \end{cases}
\end{align*}

To prove \eqref{qneq}, it remains to show the following:
Given the sequence $\{\theta(t)\}_{t=0}^{k-1}$ and time $k$, 
for any sequence of nodes 
$(m_2,\ldots,m_{\ell})\in\mathcal{V}^{\ell-1}$ with ${\ell}\leq k-1$,
there exists at most one sequence of time 
$0\leq k_1<\cdots<k_{\ell}\leq k-1$ such that
\begin{equation}
  \left[
     R_{\theta (k_{\ell})} R_{\theta (k_{{\ell}-1})} \cdots R_{\theta(k_1)}
  \right]_{ij} 
     = q_{i m_{\ell}} q_{m_{\ell} m_{{\ell}-1}}\cdots q_{m_{2}j}
\label{eqn:C_and_q}
\end{equation}
for $i\in\mathcal{V}$, where in the left-hand side, the product of $R_i$ is interpreted
as being obtained from \eqref{eqn:prod2}. 
This is shown by establishing that if \eqref{eqn:C_and_q} holds,
then the sequence of time $k_1,\ldots,k_{\ell}$ is uniquely determined
from $\{\theta(t)\}_{t=0}^{k-1}$. This can be done through the 
procedure below:
\begin{itemize}
 \item $k_1$ is the smallest $t\geq 0$ such that $\theta(t)=j$;
 \item $k_2$ is the smallest $t> k_1$ such that $\theta(t)=m_2$;
 \item $\cdots$
 \item $k_{\ell}$ is the smallest $t> k_{{\ell}-1}$ such 
       that $\theta(t)=m_{\ell}$.
\end{itemize}
We have now proven that (\ref{qneq}) holds for arbitrary $i,j,k$.

(ii)~Here, we study the average dynamics of the 
randomized update scheme
\eqref{eqn:gdra1v}. To this end, let the average matrices be
given by
$\overline{Q}:= \mathbb{E}\left[ Q_{\theta(k)} \right]$ and
$\overline{R}:=\mathbb{E}\left[ R_{\theta(k)} \right]$.
Since the updated pages are selected in an i.i.d.\ manner
from the uniform distribution, we have
\begin{equation}
  \overline{Q}
    =\frac{n-1}{n}I+\frac{1}{n}Q,~~
  \overline{R}
    =\frac{1}{n}Q.
\label{ex_a1}
\end{equation}
Here, by \eqref{eqn:xk}, 
the expectation of $x(k)$ is obtained as
\begin{align}
 &\mathbb{E}\left[ x(k) \right]
  = \mathbb{E}\biggl[ \displaystyle	
        \frac{m}{n}\mathbf{1}_n
         + \sum_{t=0}^{k-1}
            R_{\theta(t)} Q_{\theta(t-1)}\cdots Q_{\theta(0)}
           \frac{m}{n}\mathbf{1}_n 
       \biggr]	\nonumber\\
 &= \frac{m}{n}\mathbf{1}_n
     + \displaystyle	 
         \sum_{t=0}^{k-1}\mathbb{E}\left[ R_{\theta(t)}\right]
            \mathbb{E}\left[Q_{\theta(t-1)}\right]\cdots   
             \mathbb{E}\left[Q_{\theta(0)}\right]
              \frac{m}{n}\mathbf{1}_n 	\nonumber\\
  &= \frac{m}{n}\mathbf{1}_n
      + \displaystyle	
        \sum_{t=0}^{k-1}\overline{R} \overline{Q}^{t}
         \frac{m}{n}\mathbf{1}_n.
\label{tmp-a1}
\end{align}
Notice that in the second term in the far right-hand side above, 
the average link matrix $\overline{Q}$ in \eqref{ex_a1} 
is Schur stable 
since it is a nonnegative matrix whose column sums are equal to
${(n-1)}/{n} + {(1-m)}/{n} = 1-{m}/{n} < 1$.
Thus, taking the limit $k\to \infty$, 
we can apply the Neumann series as 
\[
 \displaystyle\lim_{k\to \infty}
   \sum_{t=0}^{k-1}\overline {R}\overline{Q}^{t}
    = \overline {R}
       \left(I-\overline {Q}\right)^{-1}.
\]
Moreover, by \eqref{ex_a1}, we have
\begin{align}
  &\overline{R}
     \left(
        I-\overline {Q}
     \right)^{-1} 
    \nonumber\\
  &= \overline {R}
       \biggl[
         I-\biggl( 
              \frac{n-1}{n}I+\frac{1}{n}Q 
           \biggr)
       \biggr]^{-1}
   = n\overline {R}\left(I-Q \right)^{-1}  
     \nonumber\\
  &= n\overline {R} \displaystyle
       \lim_{k\to \infty}\sum_{t=0}^{k}Q^{t} 
   = \displaystyle\lim_{k\to \infty}Q\sum_{t=0}^{k} Q^{t}
   = \displaystyle\lim_{k\to \infty}\sum_{t=1}^{k}Q^{t}.
\label{used_nounif}
\end{align}
Substituting this into (\ref{tmp-a1}) as $k\to\infty$
and by \eqref{eq_kasaneawase}, we obtain
\begin{align*}
  & \displaystyle\lim_{k\to \infty}
     \mathbb{E}\left[ x(k) \right]
  = \frac{m}{n}\mathbf{1}_n 
     + \displaystyle\lim_{k\to \infty}	
         \sum_{t=0}^{k-1}\overline {R}\overline {Q}^{t} 
          \frac{m}{n}\mathbf{1}_n	
   \nonumber\\
  &= \frac{m}{n}\mathbf{1}_n
       + \displaystyle\lim_{k\to \infty}	
          \sum_{t=1}^{k} Q^{t} \frac{m}{n}\mathbf{1}_n 
   = \sum_{t=0}^{\infty} Q^{t}\frac{m}{n} \mathbf{1}_n	
   = x^{*}.
\end{align*}

From (\ref{tmp-a1}), it is clear that 
$\mathbb{E}\left[ x(k) \right]$ can be written as a step response of
a stable linear time-invariant system. This implies that it converges
to $x^*$ exponentially fast. This completes the proof of (ii).

Finally, by property (i) above, $x(k)$ is monotonically 
nondecreasing and has an upper bound $x^*$, so $x(k)$ converges with probability 1. 
Then, due to property (ii), the convergence value for $x(k)$ is $x^*$.
\hfill\mbox{$\blacksquare$}

Theorem~\ref{unif_thm} guarantees that the proposed gossip-based 
algorithm computes the true PageRank almost surely in a fully distributed
fashion. Each page keeps track of its states $x_i(k)$ and $z_i(k)$ and
when chosen by $\theta(k)$, it transmits $z_i(k)$ to its neighboring 
pages along its hyperlinks. Such hyperlinks are clearly known to the pages and
the necessary communication is limited with only one value at a time. 
Other pages not linked by page $\theta(k)$ will simply keep their states
unchanged. 

The convergence is shown to be exponential in the mean, that is,
$\mathbb{E}[x(k)]$ goes to $x^*$ exponentially fast. 
It is interesting to note
that in \eqref{used_nounif} in the proof, we have not shown
$\sum_{t=0}^{k-1}\overline {R}\overline {Q}^{t}=
\sum_{t=1}^{k}Q^{t}$,
which would indicate that in the mean the system is the same as
the synchronous one. 
Indeed, this equality holds only in the limit as $k\to\infty$. 

Our method is based on a simple reinterpretation 
of the definition of PageRank from the systems viewpoint, and  
it seems well suited for the PageRank computation in terms of 
convergence. 
The distributed algorithms proposed in 
\cite{DaiFre:17,LagZacDab:17,YouTemQiu:17}
also have exponential convergence speeds 
(under different notions). The approaches there 
rely on techniques for distributed optimization. 
The work \cite{YouTemQiu:17} views the PageRank problem as a least-squares
problem while \cite{DaiFre:17} employs a randomized version of the
so-called matching pursuit algorithms. 
On the other hand, in \cite{LagZacDab:17},
a modified gradient-descent algorithm is constructed so that the 
states of all pages remain to have the total equal to one throughout 
its execution.

It is highlighted that our approach has an advantage 
in terms of the communication loads for each node. 
As seen earlier, in the update scheme, the nodes need 
to transmit their values only over their outgoing links,
and no further communication is necessary. 
The same type of communication scheme is adopted in 
\cite{FraIshRav:15}, which is an extension of those in
\cite{IshTem:10}, and thus the algorithms there do not
exhibit exponential convergence.
Similarly, in the algorithm of \cite{DaiFre:17}, 
the nodes utilize only the outgoing links, but there 
is a difference in that the nodes must also receive 
the values from the linked pages during the same time step,
and hence communication is always bidirectional.
Meanwhile, the knowledge of the pages connected by 
the incoming links is necessary in \cite{YouTemQiu:17}. 
The scheme in \cite{LagZacDab:17} requires communication 
along both incoming and outgoing links. 
In this respect, among the different approaches,
our Algorithm~\ref{alg:2} is superior in 
requiring the least amount of communication per update. 
Further discussions on comparisons of the methods can
be found in \cite{IshSuz:18}.

\subsection{Generalization to Non-Uniform Distributions}\label{nounif-algo}

We next generalize the gossip-type distributed algorithm 
to the case where the pages will be chosen from distributions 
not limited to the uniform one. This extension is an interesting 
feature of the proposed approach and makes the algorithm more 
suitable for its use in distributed environments. 
For example, depending on the computational and communication
resources, the pages or the servers that carry out the PageRank 
computation may adjust to update at different frequencies 
\cite{ChaHadRab:16}.

The update scheme here follows Algorithm~\ref{alg:2}.
Consider an i.i.d.\ random sequence $\{\theta(k)\}$ for 
the page selections.
Let $p_i$ be the probability of page $i$ to be chosen. 
Assume that $p_i>0$ for $i$ and $\sum_{i=1}^n p_i = 1$.
In Algorithm~\ref{alg:2}, step~2) should be replaced
with the following
\begin{enumerate}
\item[2)$'$]
Select one page $\theta(k)$ based on the distribution $p_i$:
\begin{equation}
	  \text{Prob}\bigl\{ 
			\theta(k)=i  
	             \bigr\} = p_i~~\text{for $i\in\mathcal{V}$}. 
 \label{eqn:p_i}
\end{equation}
\end{enumerate}

For this algorithm, we now state the main result.

\begin{prop}\label{prop:nounif_thm}\rm
Under Algorithm~\ref{alg:2} using step~2)$'$ introduced 
above, the PageRank vector $x^*$ is computed with
$x(k)\to x^{*}$ as $k\to \infty$ with probability $1$.
\end{prop}

\textit{Proof:}
This proposition can be established similarly to 
Theorem~\ref{unif_thm} by showing
the properties~(i) and~(ii), where 
the main difference is in (ii). 
Let the matrix $P\in\mathbb{R}^{n\times n}$ be the 
diagonal matrix 
whose $i$th diagonal entry is $p_i$, i.e.,
$P:=\text{diag}(p_1,\ldots,p_n)$.
By $p_i>0$, $P$ is nonsingular. 
Next, let the average matrices of $Q_{\theta(k)}$ and $R_{\theta(k)}$ 
be respectively
\begin{align*}
 \overline{Q}^{\prime}
   &:= \mathbb{E}\left[ Q_{\theta(k)} \right]=(I-P)+ QP,\\
 \overline{R}^{\prime}
   &:=\mathbb{E}\left[ R_{\theta(k)} \right]= QP.
\end{align*}
These matrices are nonnegative. 
Moreover, for $\overline{Q}^{\prime}$, 
the sum of its $i$th column is equal to $1-m p_i$;
this means that it has
the spectral radius $\max_i 1-m p_i <1$ and thus
is Schur stable.
Now, as in the discussion around (\ref{tmp-a1}), 
we can establish
\begin{equation*}
 \mathbb{E}\left[ x(k) \right]
  = \frac{m}{n}\mathbf{1}_n
    + \displaystyle                
      \sum_{t=0}^{k-1}{\overline{R}^{\prime}}
       \bigl({\overline{Q}^{\prime}}\bigr)^t 
      \frac{m}{n}\mathbf{1}_n.
\end{equation*}
For the summation in the right-hand side, 
take the limit $k\to \infty$ and then apply 
the Neumann series to obtain 
\begin{align*}
 &\displaystyle\lim_{k\to\infty}
  \sum_{t=0}^{k-1}
   {\overline{R}^{\prime}}
        \bigl({\overline{Q}^{\prime}}\bigr)^t
  = {\overline{R}^{\prime}}
     \bigl(I-{\overline{Q}^{\prime}}\bigr)^{-1}	
   \nonumber\\
  &~~= {\overline{R}^{\prime}}
        \left[I-\left( (I-P)+QP \right)\right]^{-1}
  = {\overline{R}^{\prime}}P^{-1}
        \left(I-Q \right)^{-1} 	
    \nonumber\\
  &~~= {\overline{R}^{\prime}}P^{-1}
       \displaystyle\lim_{k\to \infty}
       \sum_{t=0}^{k} Q^{t}
   = \displaystyle\lim_{k\to \infty}
      \left( QP \right)P^{-1}\sum_{t=0}^{k} Q^{t} 	
    \nonumber\\
  &~~= \displaystyle\lim_{k\to \infty}
         \sum_{t=1}^{k} Q^{t}.	
\end{align*}
This expression is the same as (\ref{used_nounif}). The rest of the proof
follows similarly to the proof of Theorem~\ref{unif_thm}\,(ii).
\hfill\mbox{$\blacksquare$}

This gossip-type distributed algorithm 
can be carried out 
even if the probability distribution for the page selection is not uniform.
Though other algorithms may be able to deal with non-uniform selections 
\cite{ChaHadRab:16,IshTemBai:scl12,LeiChe:15}, 
in those cases, additional computations and adjustments are often required.
In contrast, in our algorithm, no change is necessary, and 
the update scheme performed by each page is exactly the same.
We have seen that the state values increase monotonically to reach the true PageRank.
This might indicate that increasing the selection probability 
of a page with a large value may lead to faster convergence. 
We will examine this idea in the numerical example later.

\section{Generalization of the Approach}
\label{sec:generalized}

We extend the randomization-based distributed algorithms 
developed in the previous section in two directions
to enhance their convergence performance and also 
the flexibility in implementation. 

First, while Algorithm~\ref{alg:2} is restricted
to allowing only one page to initiate an update at a time,
here we realize simultaneous updates by multiple nodes 
in distributed computation. 
The other extension is to incorporate update times which
are deterministic so that no randomization is necessary. 
It will be shown that an algorithm with these novel 
features possess similar convergence properties. 

In the proposed algorithm, we denote the set of
updating pages chosen at time $k$ by $\phi(k)\subset\mathcal{V}$.
This set need not be randomly determined and
may contain arbitrary number of page indices.
We now introduce the algorithm in the following.

\begin{algo}[Distributed Algorithm with Simultaneous Updates]\label{generalized_algo}\rm
\begin{enumerate}
\item For each page $i\in\mathcal{V}$, set the initial states
as $x_{i}(0)=z_{i}(0)=m/n$. 

\item At time $k$, each page $i$ decides whether to make an update
or not. Let $\phi(k)\subset\mathcal{V}$ be the set of
indices of all pages that decided to make an update. 

\item Each page $i\in\phi(k)$ transmits its value 
$z_{i}(k)$ 
over its outgoing links to pages in $\mathcal{L}^{\text{out}}_{i}$.
\item Each page $i\in\mathcal{V}$ 
makes an update in its states to obtain 
$x_i(k+1)$ and $z_i(k+1)$ as
\begin{align}
  &x_{i}(k+1)
   = x_i(k)
       + \sum_{j\in \mathcal{L}_i^{\rm in}\cap\phi(k)}
            \frac{1-m}{n_j}z_{j}(k),
    \nonumber\\
  &z_{i}(k+1)    \nonumber\\
   &~~~= \begin{cases}
	\displaystyle 
          \sum_{j\in \mathcal{L}_i^{\rm in}\cap\phi(k)}
          \frac{1-m}{n_j}z_{j}(k)
        & \text{if $i\in\phi (k)$},\\
        \displaystyle 
          z_i(k) + \sum_{j\in \mathcal{L}_i^{\rm in}\cap\phi(k)}
                   \frac{1-m}{n_j}z_{j}(k)
        & \text{otherwise}.
     \end{cases}
\label{eqn:update:alg3}
\end{align}
\end{enumerate}
\end{algo}

This algorithm has a structure similar to that of
Algorithm~\ref{alg:2}. The communication load is
minimal since the pages that initiate updates
in step~2) transmit only their states $z_i(k)$ 
and not $x_i(k)$, and this is done over their 
outgoing hyperlinks.
In step~4), the update scheme \eqref{eqn:update:alg3} 
for the states is a generalized
version of the one in \eqref{eqn:gdra1} from 
Algorithm~\ref{alg:2}. The pages receiving
state values over their incoming edges are
characterized by having a nonempty 
set $\mathcal{L}_i^{\rm in}\cap\phi(k)$,
and only these pages make changes in their states. 

For the set $\phi\subset\mathcal{V}$ of chosen pages, 
we introduce three nonnegative 
matrices $Q_{\phi}$, $R_{\phi}$, and $S_{\phi}$ as
\begin{align}
  Q_{\phi}
    &=\begin{bmatrix} 
         q_{1}(\phi) & \cdots & q_{n}(\phi) 
      \end{bmatrix},\notag\\
  R_{\phi}
    &=\begin{bmatrix} 
         r_{1}(\phi) & \cdots & r_{n}(\phi) 
      \end{bmatrix},
   \label{eqn:alg4:QRS}\\
  S_{\phi}
    &=\begin{bmatrix} 
         s_{1}(\phi) & \cdots & s_{n}(\phi) 
      \end{bmatrix},\notag
\end{align}
where the component vectors $q_i(\phi)$, $r_i(\phi)$, and 
$s_i(\phi)$, $i=1,\ldots,n$, are given by
\begin{align*}
  q_{i}(\phi)
    &= \begin{cases}
   	 q_i & \text{if $i\in\phi$},\\
  	 e_i & \text{otherwise},
       \end{cases}~~
  r_{i}(\phi)
    = \begin{cases}
 	 q_i & \text{if $i\in\phi$},\\
  	 0_n & \text{otherwise},
	\end{cases}\\
  s_{i}(\phi)
    &= \begin{cases}
  	 0_n & \text{if $i\in\phi$},\\
  	 e_i & \text{otherwise}.
       \end{cases}
\end{align*}
We note that the following three relations hold:
\begin{align}
   R_{\phi}
     &= \sum_{i\in\phi} R_i,
   \label{eqn:alg4:relation1}\\
   Q_{\phi}
     &= R_{\phi} + S_{\phi},
   \label{eqn:alg4:relation2}\\
 R_i S_{\phi}
  &= \begin{cases}
       0 & \text{if $i\in\phi$},\\
       R_{i} & \text{otherwise}.
     \end{cases}
      \label{eqn:alg4:relation3}
\end{align}

With these matrices, we can write the update
scheme of Algorithm~\ref{generalized_algo}
in a vector form as
\begin{equation}
\begin{split}
  x(k+1)
    &= x(k) + R_{\phi(k)}z(k),\\
  z(k+1)
    &= Q_{\phi(k)}z(k),
\end{split}
\label{general_q}
\end{equation}
where the initial states are $z(0)=x(0)=(m/n)\mathbf{1}_n$.

Regarding the choice of the sequence $\{\phi(k)\}$,
we make the following assumption. 
It says that each
page must initiate the updates of its states 
infinitely often over time. 

\begin{asm}\label{generalized_asm}\rm
Each page $i\in\mathcal{V}$ is contained in infinitely 
many sets $\phi(0),\phi(1),\ldots,\phi(k),\ldots$.
\end{asm}

We are now in the position to state the 
main result for the distributed algorithm 
with multiple updates.

\begin{thm}\label{geranalized_conv}\rm 
Under Assumption~\ref{generalized_asm},
in the distributed algorithm with simultaneous updates of
Algorithm~\ref{generalized_algo}, 
the state $x(k)$ converges to the true PageRank vector $x^*$,
that is, $x(k)\to x^*$ as $k\to\infty$.
If, in addition, for some $T>0$, each page updates at least once 
in every $T$ steps, then the convergence to $x^*$ 
is exponential.
\end{thm}

The proof of this theorem is a generalization
of that for Theorem~\ref{unif_thm} for Algorithm~\ref{alg:2}. 
It consists of showing two properties similar to (i) and (ii) 
in Theorem~\ref{unif_thm}. Here, we state the first property
as a lemma. Its proof is given 
in Appendix~\ref{appendix:1}, which follows similar lines
as that of Theorem~\ref{unif_thm}, but is more technical 
and involved.

\begin{lem}\label{generalized_monot}\rm
Under Assumption~\ref{generalized_asm},
in the distributed algorithm with simultaneous updates of
Algorithm~\ref{generalized_algo}, it holds
$x(k)\leq x(k+1)\leq x^*$ for $k\in\mathbb{Z}_+$.
\end{lem}

\textit{Proof of Theorem~\ref{geranalized_conv}}:
By Lemma~\ref{generalized_monot}, the state $x(k)$ of 
Algorithm~\ref{generalized_algo} converges to 
some vector $x'\leq x^*$. We must show that this $x'$ is always
equal to $x^*$. Take an arbitrary $\varepsilon>0$.
Let $k_{\text{s}}(\varepsilon)$ be the step number $k\geq 1$ such that
under the synchronous update scheme \eqref{eqn:synch}, whose
states are denoted by $x_{\text{s}}(k)$
and $z_{\text{s}}(k)$, the error bound of 
\[
   \|x_{\text{s}}(k)-x^*\|_1\leq\varepsilon
\]
is achieved for the first time. By Lemma~\ref{lemm:synch}, 
in the synchronous algorithm, the error $x_{\text{s}}(k)-x^*$ 
asymptotically converges to zero.
Thus, for any given $\varepsilon$, a finite value
for $k_{\text{s}}$ always exists.

Next, for a given sequence $\{\phi(k)\}_{k=0}^{\infty}$,
we recursively define the time sequence 
$\{\delta(k)\}_{k=0}^{\infty}$ by $\delta(0)=-1$ and 
\begin{equation}
  \delta(k)
   =\min\bigg\{
        \ell> \delta(k-1):\
         \bigcup_{t=\delta(k-1)+1}^{\ell} \phi(t) 
           = \mathcal{V} 
        \bigg\}
\label{eqn:delta}
\end{equation}
for $k\geq 1$.
By definition,
during the time interval from $\delta(k-1)$ to $\delta(k)$,
all nodes $1,2,\ldots,n$ are chosen at least once.
By Assumption~\ref{generalized_asm}, 
this sequence $\{\delta(k)\}_{k=0}^{\infty}$ is well defined.

To establish that the state $x(k)$ converges to $x^*$, 
we must show that for arbitrary $\varepsilon$, the following
inequality holds:
\begin{align}
  x^* 
   &> x(\delta(k_{\text{s}}(\varepsilon)))
   \geq x_{\text{s}}(k_{\text{s}}(\varepsilon)).
\label{eqn:thm2:1}
\end{align}
This relation indicates that as the state $x_{\text{s}}(k)$ 
of the synchronous algorithm converges to the PageRank
vector $x^*$, it will be followed by 
the state $x(k)$ of Algorithm~\ref{generalized_algo}
at a slower speed governed by $\delta(k)$.

In what follows, we prove the relation 
\begin{equation}
 \sum_{t=0}^{\delta(k_s(\varepsilon))}
    \left[\widetilde Q_{\phi}(t)  \right]_{ij}
 \geq \sum_{t=0}^{k_s(\varepsilon)}
    \left[\widetilde Q_{s}(t)  \right]_{ij}~~
   \text{for $i,j$}.
\label{general_lower}
\end{equation}
Note that this relation is similar to \eqref{general_q_neq}
in the proof of Lemma~\ref{generalized_monot},
but with the difference in the direction of the inequality
and the times over which the summations are taken. 
By \eqref{eqn:Qs_ij} of Lemma~\ref{lem:synch}, 
we can rewrite the right-hand side 
of \eqref{general_lower} as
\begin{equation*}
 \sum_{t=0}^{k_s(\varepsilon)}
   \left[ \widetilde{Q}_{\rm{s}}(t) \right]_{ij}
    = \sum_{t=0}^{k_s(\varepsilon)}
        \sum_{m_1,\ldots,m_t\in \mathcal V}
          q_{im_t}q_{m_tm_{t-1}} \cdots q_{m_1j}.
\end{equation*}

We now focus on the left-hand side of \eqref{general_lower}.
Note that similarly to the proof of Lemma~\ref{generalized_monot},
it is the sum of the terms appearing in \eqref{general_meidai}
at most once.
For each $t\leq \{0,1,\ldots,k_s(\varepsilon)\}$, consider the ordered set of nodes,
$(m_1,m_2,\ldots,m_t)\in \mathcal{V}^{t}$.
We show that for each $i,j$, the term 
$q_{im_t}q_{m_tm_{t-1}}\cdots q_{m_1j}$ 
corresponding to this set of nodes
appears once in the left-hand side
of \eqref{general_lower}.

Consider the sequence $\left( k_{t+1},k_{t},\cdots,k_1 \right)$
of time, where
\begin{itemize}
 \item $k_1$ is the smallest $\ell\geq0$ such that $m_1\in\phi(\ell)$; 
 \item $k_2$ is the smallest $\ell>k_1$ such that $m_2\in\phi(\ell)$;
 \item $\ldots$
 \item $k_t$ is the smallest $\ell>k_{t-1}$ such that $m_t\in\phi(k)$.
\end{itemize}
The sequence $\left( k_{t+1},k_{t},\ldots,k_1 \right)$ exists.
This is because by the choice of $\delta(\cdot)$
in \eqref{eqn:delta}, it holds $\delta(t-1)<k_t\leq \delta(t)$. 
Therefore, it follows that the left-hand side 
of \eqref{general_lower}
is a sum of nonnegative terms, and
moreover, it contains all terms appearing in the right-hand side.
We therefore conclude that the relations \eqref{general_lower} 
and thus \eqref{eqn:thm2:1} hold.

Since this holds for arbitrary $\varepsilon$,
it finally follows that $x(k)$ converges to $x^*$.
Under the additional assumption that each page
updates at least once in every $T$ steps, 
it holds $\delta(k+1)-\delta(k)\leq T$. 
Since $x_{\text{s}}(k)$ exponentially converges to $x^*$,
$x(k)$ does as well. 
\hfill\mbox{$\blacksquare$}

\section{Clustering-Based Distributed Algorithm}
\label{sec:group}

In this section, we develop a novel approach based
on web clustering
for the computation of PageRank.

In this context, web clustering means the following:
(i)~Prior to running the algorithm, 
we group the pages, preferably, having strong dependence
on each other through hyperlinks.
(ii)~During the computation, we allow the pages 
within groups to communicate with each other 
for updating their states together. 
That is, in this case, the states are updated not by 
the individual pages but by the groups. In doing so,
we assume that extra computation resources are available
locally within the group, which will be exploited to 
expedite the convergence speed. 
This scheme is especially suitable in view of the structure of
the web since pages belong to
domains and subdomains, which can be directly 
considered as groups.

Even through the grouping,
our approach is able to compute the true PageRank
values and, more important, this can be realized
much more efficiently in terms of computation speed.
This advantage is realized by introducing extra local
computation in the group-wise updates, where
multiplications based on submatrices of the link matrices
are performed. In previous research, grouping of pages
often arises as part of an aggregation process, where
PageRank values representing the groups are computed
for reducing the size of the problem;
see, e.g., 
\cite{BroLem_infret:06,IshTemBai:tac12,LanMey:06,ZhuYeLi:05}.

\subsection{Grouping of Pages}

We partition the web consisting of $n$ pages into $N\leq n$ groups, 
denoted by 
$\mathcal{V}_{1},\mathcal{V}_{2},\ldots,\mathcal{V}_{N}\subset\mathcal{V}$.
Here, let $l_h:=|\mathcal{V}_{h}|$ be the size of group $h=1,2,\dots,N$.
Then, the constraints on the grouping are
$l_h \geq 1$, $\bigcup_{h=1}^{N}\mathcal{V}_{h}=\mathcal{V}$,
and $\mathcal{V}_{h_1}\cap\mathcal{V}_{h_2}=\emptyset$ for $h_1\neq h_2$.
It is however expected that the convergence of the computation would be faster
if groups are dense in the links among the group member pages
and if the numbers of links going outside the groups are small.  

The order of the indices can be changed 
without loss of generality and is done
according to the grouping as follows:
\begin{align*}
  \mathcal{V}_{1}&=\left\{ 1,2,\ldots,l_1 \right\},\\
  \mathcal{V}_{2}&=\left\{ l_1+1,l_1+2,\ldots,l_1+l_2 \right\},\\
  &\vdots\\
  \mathcal{V}_{N}&=\left\{ n-l_N+1,n-l_N+2,\ldots,n \right\}.
\end{align*}
After this renaming of the pages, let $A$ be 
the link matrix of the web. 
Recall that $Q=(1-m)A$. We partition this matrix $Q$ 
according to the groups as
\begin{equation}
  Q = \begin{bmatrix}
 	\check{Q}_{11} & \check{Q}_{12} & \cdots & \check{Q}_{1N} \\
	\check{Q}_{21} & \check{Q}_{22} & \cdots & \check{Q}_{2N} \\
	\vdots & \vdots & \ddots & \vdots \\
	\check{Q}_{N1} & \check{Q}_{N2} & \cdots & \check{Q}_{NN}
      \end{bmatrix},
\label{eqn:Qpartition}
\end{equation}
with the submatrices
$\check{Q}_{ij}\in\mathbb{R}^{l_{i}\times l_{j}}$, 
$i,j=1,2,\ldots,N$.
Similarly, the states $x(k)$ and $z(k)$ are partitioned as
\begin{equation}
\begin{split}
  x(k) &= \begin{bmatrix} 
           \check{x}_{1}(k)^T & \check{x}_{2}(k)^T & 
              \cdots & \check{x}_{N}(k)^T 
         \end{bmatrix}^{T},\\
  z(k) &= \begin{bmatrix} 
           \check{z}_{1}(k)^T & \check{z}_{2}(k)^T & 
              \cdots & \check{z}_{N}(k)^T 
         \end{bmatrix}^{T},
\end{split}
\label{eqn:xz_group1}
\end{equation}
where 
$\check{x}_{h}(k),\check{z}_{h}(k)\in\mathbb{R}^{l_h}$
denote the states for group $h$.


In the clustering-based algorithm to be presented, one group at each
time makes an update. The group making updates at time $k$ 
is denoted by $\psi(k)\in\{1,2,\ldots,N\}$. In Algorithm~\ref{generalized_algo} 
from Section~\ref{sec:generalized}, this can be expressed as 
$\phi(k)=\mathcal{V}_{\psi(k)}$ at each time $k$. 
In the next subsection, we introduce a novel method for 
accelerating the convergence in the state updates.

\subsection{Group-Based Update Scheme}


The idea behind our approach for clustering-based computation
is that at time $k$, the pages within the chosen 
group $\psi(k)$ make updates at once 
as in Algorithm~\ref{generalized_algo}. 
The major difference however is that when they do so, 
they make a large number of updates
by exploiting the information locally available 
within the group. 
More concretely, at time $k$, 
based on their present states $\check{x}_{\psi(k)}(k)$ and
$\check{z}_{\psi(k)}(k)$, the members of
the group~$\psi(k)$ update their states infinitely 
many times. Then, the asymptotic values will be set 
as the next states $\check{x}_{\psi(k)}(k+1)$ and
$\check{z}_{\psi(k)}(k+1)$. Note that the
infinite updates are based only on intra-group 
communications, and those with pages in other groups will be performed
only after the updates in the group are completed.
We show that 
the infinite updates within the group can be 
done in one step, by small-scale matrix operations.

We first derive the update scheme for the states. 
At time $k$, assume that group $\psi(k)$ is
chosen to make updates for the rest of the time $t\geq k$. 
We write the update scheme in \eqref{general_q} 
with auxiliary states $x'(t|k)$ and $z'(t|k)$
whose initial values are set as
$x'(k|k)=x(k)$ and $z'(k|k)=z(k)$.
For time $t\geq k$, it holds
\begin{align*}
  x'(t+1|k)
    &= x'(t|k) + R_{\psi(k)} z'(t|k),\\
  z'(t+1|k)
    &= Q_{\psi(k)} z'(t|k).
\end{align*}
By using the partition of $Q$ in \eqref{eqn:Qpartition},
and also by partitioning $x'(t|k)$ and $z'(t|k)$ as in 
\eqref{eqn:xz_group1}, 
the states in the updating group 
$\psi(k)$ are given as
\begin{equation}
\begin{split}
 \check{x}'_{\psi(k)}(t+1|k)
  &= \check{x}'_{\psi(k)}(t|k)
     + \check{Q}_{\psi(k)\psi(k)} 
        \check{z}'_{\psi(k)}(t|k),\\
 \check{z}'_{\psi(k)}(t+1|k)
  &= \check{Q}_{\psi(k)\psi(k)} 
       \check{z}'_{\psi(k)}(t|k).
\end{split}
\label{eqn:xz_psi1}
\end{equation}
Moreover, the states in any remaining group $h\neq\psi(k)$ 
are updated by
\begin{equation}
\begin{split}
 \check{x}'_{h}(t+1|k)
  &= \check{x}'_{h}(t|k)
     + \check{Q}_{h \psi(k)}
         \check{z}'_{\psi(k)}(t|k),\\
 \check{z}'_{h}(t+1|k)
  &= \check{z}'_{h}(t|k)
      + \check{Q}_{h \psi(k)}
          \check{z}'_{\psi(k)}(t|k).
\end{split}
\label{eqn:xz_h1}
\end{equation}

As mentioned above, in this algorithm, the states 
$x(k+1)$ and $z(k+1)$ are taken as the limits of
the states $x'(t|k)$ and $z'(t|k)$ with $t\to\infty$
while the same group $\psi(k)$ is continuously 
chosen infinitely many times. 
First, the state $\check{z}_{\psi(k)}(k+1)$ 
of the chosen group $\psi(k)$
is set to zero because by \eqref{eqn:xz_psi1},
we have
\begin{align}
 \check{z}_{\psi(k)}(k+1)
   &= \lim_{t\to\infty}
       \check{z}'_{\psi(k)}(t|k)\notag\\
   &= \lim_{t\to\infty}
       \check{Q}_{\psi(k)\psi(k)}^{t-k}
         \check{z}_{\psi(k)}(k|k)
   = 0.
\label{eqn:z_V_k1}
\end{align}
Note that the submatrix $\check{Q}_{\psi(k)\psi(k)}$ 
is Schur stable since it is a submatrix of $Q$, 
which is nonnegative and Schur stable.

Second, from \eqref{eqn:xz_h1}, 
the state $\check{z}_{h}(k+1)$ 
of group $h\neq \psi(k)$
can be obtained by using the relation 
in \eqref{eqn:z_V_k1}
and by the Schur stability of the matrix $\check{Q}_{\psi(k)\psi(k)}$
as follows:
\begin{align}
 &\check{z}_{h}(k+1)
 = \lim_{t\to\infty}
     \check{z}'_{h}(t|k)\notag\\
 &~= \check{z}'_{h}(k|k)
     + \lim_{t\to\infty}
       \check{Q}_{h \psi(k)}
       \sum_{l=k}^{t-1}
         \check{z}'_{\psi(k)}(l|k)
  \notag\\
 &~= \check{z}'_{h}(k|k)
     + \lim_{t\to\infty}
     \check{Q}_{h \psi(k)}
       \sum_{l=0}^{t-k-1}
            \check{Q}^{l}_{\psi(k)\psi(k)}
           \check{z}_{\psi(k)}(k|k)\notag\\
 &~= \check{z}_{h}(k)
    + \check{Q}_{h \psi(k)}
          \Big( 
             I - \check{Q}_{\psi(k)\psi(k)} 
          \Big)^{-1} 
           \check{z}_{\psi(k)}(k),
  \label{eqn:z_Vh2}
\end{align}
where in the last equality, the Neumann series is used. 

Third, observe in \eqref{eqn:xz_psi1} and \eqref{eqn:xz_h1}
that the state $\check{x}_{h}(k+1)$ 
takes similar forms for both $h=\psi(k)$ and
$h\neq\psi(k)$. Thus, by a derivation similar
to \eqref{eqn:z_Vh2}, we have
\begin{align}
 &\check{x}_{h}(k+1)
  = \lim_{t\to\infty}
       \check{x}'_{h}(t|k)
   \notag\\
 &~= \check{x}'_{h}(k|k)
       + \lim_{t\to\infty} 
           \check{Q}_{h\psi(k)}
          \sum_{l=k}^{t-1}
           \check{z}'_{\psi(k)}(l|k)\notag\\
 &~= \check{x}_{h}(k) 
       + \check{Q}_{h\psi(k)} 
          \Big( 
             I - \check{Q}_{\psi(k)\psi(k)} 
          \Big)^{-1} 
          \check{z}_{\psi(k)}(k).
\label{eqn:x_Vpsi2}
\end{align}

To summarize the discussion above, 
by \eqref{eqn:z_V_k1}--\eqref{eqn:x_Vpsi2},
we arrive at the following distributed algorithm 
for the PageRank computation
based on web clustering.

\begin{algo}[Clustering-Based Distributed Algorithm]
\label{alg:aggregation}\rm
\begin{enumerate}
\item For each group $h\in\{1,2,\ldots,N\}$, 
set the initial values of the states 
as $\check{x}_{h}(0)=\check{z}_{h}(0)=m/n\mathbf{1}_{l_h}$. 

\item At time $k$, one group $\psi(k)\in\{1,2,\ldots,N\}$ 
is chosen for making updates in the states.

\item 
Using its own state $\check{z}_{\psi(k)}(k)$,
group $\psi(k)$ computes the auxiliary vector
\begin{equation}
  \overline{\check{z}}_{\psi(k)}(k)
   :=             \big( 
               I-\check{Q}_{\psi(k)\psi(k)} 
             \big)^{-1}
           \check{z}_{\psi(k)}(k)
\label{eqn:z_check_bar}
\end{equation}
and transmits it over outgoing links to groups 
containing pages 
in $\mathcal{L}^{\text{out}}_{i}$ for $i\in\psi(k)$.

\item Each group $h$ updates its states to obtain
$\check{x}_{h}(k+1)$ and
$\check{z}_{h}(k+1)$ as follows:
\begin{equation}
\begin{split}
  \check{x}_{h}(k+1)
  &= \check{x}_{h}(k) 
       + \check{Q}_{h\psi(k)}
           \overline{\check{z}}_{\psi(k)}(k),\\
  \check{z}_{h}(k+1)
  &= \begin{cases}
       0 &  \text{if $h=\psi (k)$},\\
       \check{z}_{h}(k) 
        + \check{Q}_{h\psi(k)}
           \overline{\check{z}}_{\psi(k)}(k)
       & \text{otherwise}.
      \end{cases}
\end{split}
\label{eqn:alg:aggregation}
\end{equation}
\end{enumerate}
\end{algo}

We interpret the updates made by one group $h=\mathcal{V}_{\psi(k)}$
at each time $k$ to be those made by the
member pages in the group simultaneously. 
In this way, the argument in Section~\ref{sec:generalized}
can be similarly applied to the clustering-based case. 

This clustering-based algorithm has several
advantageous features in terms of computation
speed and distributed implementation. 
To be more specific, in this algorithm, one update 
by a group~$h$
corresponds to an infinite number of updates
by the pages in the group in Algorithm~\ref{generalized_algo}.
As seen in \eqref{eqn:z_check_bar},
it involves matrix operations of the size of
the group. This can greatly accelerate the 
convergence in one step compared to the previous
algorithms. The performance would likely improve
especially by grouping the pages so that more 
groups consist of dense subgraphs
in the web, and/or each group has a limited number 
of links going outside.

Furthermore, in comparison to 
Algorithm~\ref{generalized_algo}, 
the main additional computation 
in the iteration is 
step 3) for obtaining 
$\overline{\check{z}}_{\psi(k)}(k)$
in \eqref{eqn:z_check_bar}. 
Note however that this is done locally
within the group, and 
$\overline{\check{z}}_{\psi(k)}(k)$ need not be
stored for the next step.
Also, 
the matrix inversion for computing
$\left( I-\check{Q}_{hh} \right)^{-1}$ only once
within each group~$h$
can be performed offline prior
to running the algorithm. 
We should note that the matrix $\check{Q}_{hh}$ may be
a sparse matrix in general for PageRank, but
the matrix $\left( I-\check{Q}_{hh} \right)^{-1}$ may 
have a dense structure. 

We express the algorithm above in a vector form.
It is useful to notice
that by the definition \eqref{eqn:alg4:QRS} of 
$R_{\mathcal{V}_{\psi}(k)}$, we have
\begin{align*}
  R_{\mathcal{V}_{\psi}(k)}^{t+1}
  &= \begin{bmatrix}
      0 & \check{Q}_{1\psi(k)} 
          \check{Q}_{\psi(k)\psi(k)}^{t} & 0\\
      0 & \check{Q}_{2{\psi}(k)} 
          \check{Q}_{\psi(k)\psi(k)}^{t} & 0\\
      \vdots & \vdots & \vdots\\
      0 & \check{Q}_{N\psi(k)} 
          \check{Q}_{\psi(k)\psi(k)}^{t} & 0
    \end{bmatrix}
\end{align*}
for $t\geq 0$, 
where only the columns corresponding to
the chosen group $\mathcal{V}_{\psi}(k)$ are nonzero.
Due to the matrix
$\check{Q}_{{\psi}(k){\psi}(k)}$ being
Schur stable,
by applying the Neumann series, we obtain
\begin{align}
 \sum_{t=0}^{\infty}
  R_{\mathcal{V}_{\psi}(k)}^{t+1}
  &= \begin{bmatrix}
      0 & \check{Q}_{1{\psi}(k)} 
          \big(
            I-\check{Q}_{{\psi}(k){\psi}(k)}
          \big)^{-1}
        & 0\\
      0 & \check{Q}_{2{\psi}(k)} 
          \big(
            I - \check{Q}_{{\psi}(k){\psi}(k)}
          \big)^{-1}
        & 0\\
      \vdots & \vdots & \vdots\\
      0 & \check{Q}_{N{\psi}(k)} 
          \big(
            I-\check{Q}_{\psi(k)\psi(k)}
          \big)^{-1}
        & 0
    \end{bmatrix}.
\label{eqn:Q_h_psi2}
\end{align}
Denote this matrix by $\widehat{R}_{\mathcal{V}_{\psi}(k)}$.
Since the matrix $R_{\mathcal{V}_{\psi}(k)}$ is
also Schur stable, it follows that 
\begin{align}
 \widehat{R}_{\mathcal{V}_{\psi}(k)}
  :=\sum_{t=0}^{\infty}
      R_{\mathcal{V}_{\psi}(k)}^{t+1}
  = R_{\mathcal{V}_{\psi}(k)}
      \big(
         I - R_{\mathcal{V}_{\psi}(k)}
      \big)^{-1}.
\label{eqn:Q_h_psi3}
\end{align}
Now, by \eqref{eqn:Q_h_psi2} and \eqref{eqn:Q_h_psi3},
the updates of $x(k)$ in \eqref{eqn:alg:aggregation}
of Algorithm~\ref{alg:aggregation} can be 
written as
\begin{align*}
 x(k+1)
  &= x(k) 
    + 
      \widehat{R}_{\mathcal{V}_{\psi}(k)}
       z(k).
\end{align*}
Further, based on \eqref{eqn:Q_h_psi3} and the definition
of $S_{\mathcal{V}_{\psi(k)}}$ in \eqref{eqn:alg4:QRS}, 
we can write 
the updates of $z(k)$ in \eqref{eqn:alg:aggregation}
as follows: 
\begin{align*}
 z(k+1)
  &= S_{\mathcal{V}_{\psi(k)}}
      \big(
        I + \widehat{R}_{\mathcal{V}_{\psi}(k)}
      \big)z(k).
\end{align*}
The initial states are set as $x(0)=z(0)=(m/n)\mathbf{1}_n$.

We introduce an assumption regarding $\psi(k)$.

\begin{asm}\label{aggregation_asm}\rm
Each group $i$ is chosen infinitely many
times in $\psi(0),\psi(1),\ldots,\psi(k),\ldots$.
\end{asm}

The main result for this algorithm is stated as follows:

\begin{prop}\label{aggregation_conv}\rm 
Under Assumption~\ref{aggregation_asm},
in Algorithm~\ref{alg:aggregation},
the state $x(k)$ converges to the true PageRank vector $x^*$, 
that is, $x(k)\to x^*$ as $k\to\infty$.
If, in addition, for some $T>0$, each group is chosen at least once 
in every $T$ steps, then the convergence to $x^*$ 
is exponential.
\end{prop}

\begin{figure}[tb]
        \vspace*{2mm}
	\centering
        \hspace*{-2mm}
	\includegraphics[width=9cm]{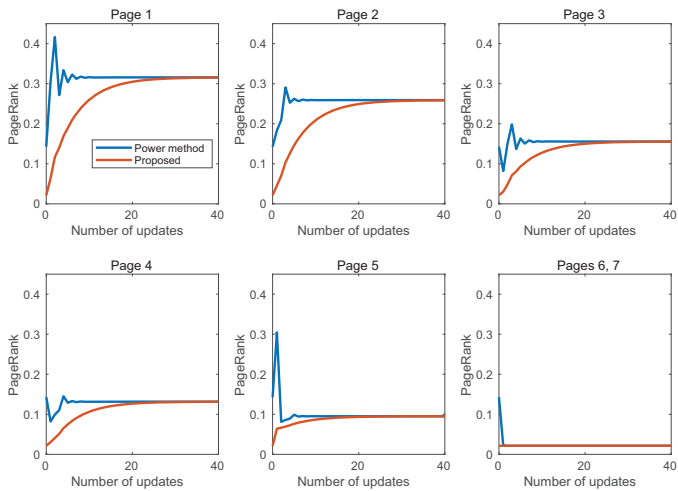}
        \vspace*{-5mm}
	\caption{Time responses of the synchronous algorithms for the small graph: 
        The power method and Algorithm~\ref{alg:1}}
	\label{simple-cent}
\end{figure}

The proof of this proposition relies
on the following lemma,
whose proof is presented in Appendix~\ref{appendix:2}.   

\begin{lem}\label{aggregation_monot}\rm
In the distributed algorithm based on clustering of
Algorithm~\ref{alg:aggregation}, it holds
$x(k)\leq x(k+1)\leq x^*$ for $k\in\mathbb{Z}_+$.
\end{lem}

With this lemma, 
the proof of Proposition~\ref{aggregation_conv}
itself follows similarly to that of 
Theorem~\ref{geranalized_conv} and is
hence omitted. 
The crucial difference however is due to 
the infinite updates made within groups 
in Algorithm~\ref{alg:aggregation}.
This aspect becomes evident by comparing 
Lemma~\ref{generalized_monot} (for Theorem~\ref{geranalized_conv})
and Lemma~\ref{aggregation_monot} (for Proposition~\ref{aggregation_conv}).
Specifically, 
the finite summation in \eqref{general_q_neq1} in 
the proof of the former result will be replaced with
an infinite one, which is found in \eqref{aggr_proof_1}
in the proof of the latter. 




\section{Numerical Examples}\label{sec:example}

In this section, we illustrate the performance of the 
proposed algorithms by numerical simulations
and compare them with conventional methods. 
Our update schemes are applied to two graphs,
a simple one and one from actual web data. 

\subsection{Small Graph}\label{case-simple}

We first use the simple graph with seven pages in Fig.~\ref{example1}.

\begin{figure}[t]
        \vspace*{3mm}
	\centering
        \hspace*{-1.5mm}
	\includegraphics[width=9cm]{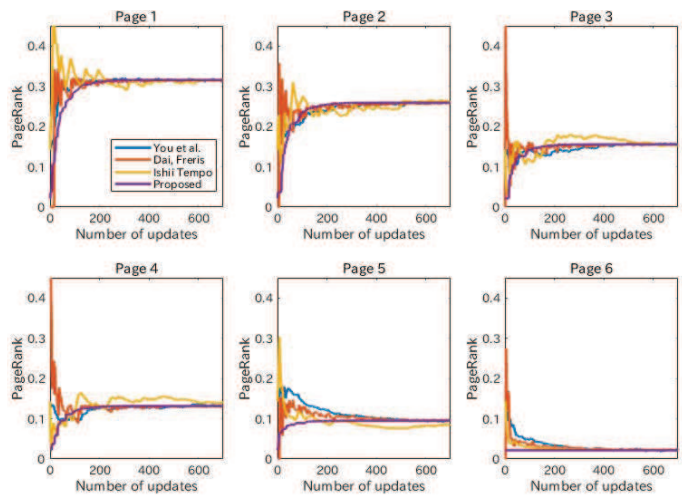}
        \vspace*{-5mm}
	\caption{Time responses of the asynchronous algorithms for the small graph: 
        You et al.~\cite{YouTemQiu:17}, Dai and Freris~\cite{DaiFre:17}, 
        Ishii and Tempo~\cite{IshTem:10},
        and Algorithm~\ref{alg:2}}
	\label{simple-dist}
\end{figure}

\subsubsection{Synchronous Algorithms}

We compare two synchronous algorithms, the power method and our proposed
Algorithm~\ref{alg:1}.
The two algorithms differ in their initial states. 
The proposed algorithm requires $x(0)$ to be $(m/n)\mathbf{1}_n$ while
the power method can take any initial state as long as it is a probabilistic vector;
in this simulation for the latter, we also used uniform values, i.e., 
$(1/n)\mathbf{1}_n$.
On the other hand, these two algorithms are both 
deterministic and, as a consequence, 
the responses of pages~$6$ and $7$ become exactly the same. 


The time responses of the PageRank values for the seven
pages are shown in Fig.~\ref{simple-cent}.
We observe that the power method converges faster in most nodes.
In the responses of the proposed algorithm, it is noticeable that they are 
nondecreasing with respect to time, a property shown in Lemma~\ref{lemm:synch}\,(i).
Also, recall that for pages~6 and 7, in the proposed 
algorithm, the PageRank values 
are equal to the assigned initial values $m/n$. Hence,
for these pages, the proposed algorithm is faster.

\subsubsection{Gossip-Type Distributed Algorithms}\label{randomized_compare}

Next, we discuss the results for the proposed 
randomized distributed algorithm, 
Algorithm~\ref{alg:2}, based on 
the uniform distribution for the gossip-based
communication. 
We make comparisons of its performance with several randomized algorithms in the literature.
Specifically, we implemented those of
Ishii and Tempo \cite{IshTem:14}, 
Dai and Freris \cite{DaiFre:17}, 
and You, Tempo, and Qiu \cite{YouTemQiu:17}.
In the latter two algorithms, the total number $n$ of pages in the web
is considered unknown though it is of course needed for the calculation 
of the PageRank as defined in \eqref{def-normal}; here, we assume that $n$ is known
by all pages. 

All four algorithms select one page at each time $k$ based on the uniform distribution,
and we used the same sequence $\{\theta(k)\}$.
Concerning the initial states, our proposed algorithm 
and that of \cite{DaiFre:17}
require that the pages take a fixed value, respectively,
equal to $m/n$ and $0$.
Other algorithms have some freedom in the choices. 
Here, however, we set them so that all pages are 
given the same initial values and took $1/n$.

The time responses of the calculated PageRank values of the nodes are plotted in 
Fig.~\ref{simple-dist}.
We omit the result for page~$7$ as its behavior is similar to that of page~6.
It is clear that the proposed algorithm is the fastest
in terms of convergence time for all pages in comparison with other distributed randomized algorithms. 
The responses of the proposed algorithm are characteristic in that despite 
the randomization due to gossipping,
they are very smooth and again nondecreasing as in Fig.~\ref{simple-cent}. 


\begin{figure}[t]
  \centering
  \includegraphics*[width=8cm]{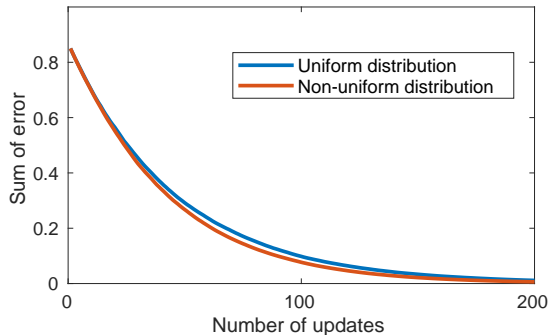}
  \vspace*{-1mm}
  \caption{Time responses of the sums of errors
for Algorithm~\ref{alg:2} in the small graph:
Comparison between two distributions,
          uniform and non-uniform}        
\label{nounif-dist}
\end{figure}


\subsubsection{Comparison of Distributions in Page Selection}

Here, we illustrate how the convergence speed can be improved
by employing Algorithm~\ref{alg:2} with a non-uniform 
distribution \eqref{eqn:p_i} for $\theta(k)$.
As discussed earlier,
it seems reasonable to increase the selection probability of pages expected to take 
larger PageRank values.
We adjusted the probabilities so that 
pages having more incoming links are more likely to be selected.
In particular, we assigned each page the probability proportional to
its in-degree plus 1.

We made Monte Carlo simulations of 1,000 runs
by executing Algorithm~\ref{alg:2} for two cases:
One under the uniform distribution
and the other under the non-uniform distribution.
The time responses of the sample averages
of the sum of the errors, i.e.,
$\|x(k)-x^*\|_1$,
are shown in Fig.~\ref{nounif-dist}.
The non-uniform distribution slightly accelerates the convergence.
It remains to be investigated what kind 
of distribution can be beneficial.


\begin{figure}[tb]
	\centering
        \hspace*{0cm}
        \includegraphics[width=9.9cm]{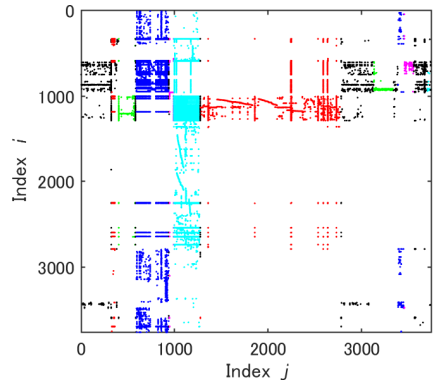}
        \vspace*{-5mm}
	\caption{Web graph of the large network}
	\label{fig:Lincoln_graph}
\end{figure}

\begin{figure}[tb]
	\centering
        \includegraphics[width=8.4cm]{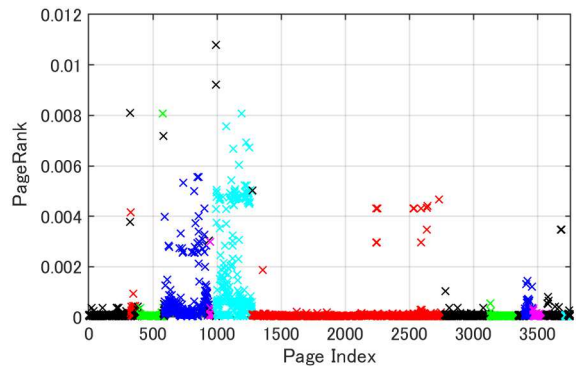}
        \vspace*{-4mm}
	\caption{PageRank values of the pages in the large network:
                 Markers are colored for groups larger than 20 nodes.}
	\label{fig:PageRank_agg}
\end{figure}

\subsection{Clustering-Based Algorithm Using Real Web Data}

In this subsection, we apply the proposed algorithms 
including the one based on clustering
to real data of the web and demonstrate their
performance. 

\subsubsection{Web Data and Clustering}
As the web graph, we used data 
from the database \cite{StaCybRes:06}
collected from Lincoln University in New Zealand in 2006. 
This data has been used as a benchmark for testing 
different algorithms related to PageRank 
in, e.g., \cite{FerAkiBou:tac12,IshTem:14,YouTemQiu:17}. 
In particular, we adopted the data from \cite{IshTem:14},
which is slightly modified to remove unlinked pages
and to add additional links for dangling nodes. 
It consists of $n=3,754$ pages with 40,646 hyperlinks.
Further details regarding the data can be found
in \cite{IshTem:14}.

We ordered the pages alphabetically according to their 
address names and then grouped them. First, for the 2,891 
pages in the university domain (www.lincoln.ac.nz), 
they were grouped based on the first subdomain names. 
Then, the remaining
pages outside the university were 
grouped based on their domain names. 
In total, the number of groups is 718, 
numerically indexed starting with group~1
containing page~1.

The larger groups contain 1,502, 346, 282, 221 pages
and so on, but then there were 594 groups 
with only one page and 78 groups with two pages. 
The graph structure is shown in Fig.~\ref{fig:Lincoln_graph},
where the dots indicate the nonzero entries of
the hyperlink matrix $A$. 
Here, the colored columns correspond to some of the 
larger groups, containing more than 20 member pages;
there are 10 such groups. 
Also, in Fig.~\ref{fig:PageRank_agg}, the PageRank 
values of the pages are plotted, with the same 
coloring scheme. 

Observe that the pages of the group colored in
blue (group~282, around page index 700, 
with 346 pages) and those in the
group shown in cyan color (group~290, around page index 1,100, with 282 pages) 
take especially high PageRank values.
As seen in Fig.~\ref{fig:Lincoln_graph},
these groups have fairly dense
link structures within their groups
and many incoming links from outside.
The pages taking the two highest PageRank values 
(page indices 991 and 992) are
the university search page, which form a group 
by themselves; they receive about 270 incoming links.

\begin{figure}[t]
	\centering
         \includegraphics[width=8.3cm]{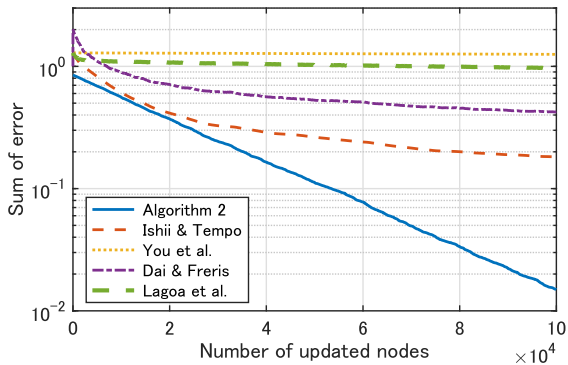}
        \vspace*{-1mm}
	\caption{Time responses of 
        errors in asynchronous randomized algorithms:
        Algorithm~\ref{alg:2},
        Ishii and Tempo~\cite{IshTem:10},
        You et al.~\cite{YouTemQiu:17}, 
        Dai and Freris~\cite{DaiFre:17}, 
        and Lagoa et al.~\cite{LagZacDab:17}}
	\label{fig:single_update}
\end{figure}

\subsubsection{Comparison among Distributed Algorithms}
We discuss the results for the 
asynchronous distributed algorithm, Algorithm~\ref{alg:2}.
As in the previous subsection,
we make comparisons of its performance with 
the algorithms from 
\cite{DaiFre:17}, \cite{IshTem:14}, 
\cite{LagZacDab:17}, and \cite{YouTemQiu:17},
which are all randomization based.
All five algorithms select one page at each time $k$ 
based on the uniform distribution,
and we used the same sequence $\{\theta(k)\}$.
As initial states, all pages were given the
same values. In our algorithms, 
this is $m/n$. For the algorithm of \cite{DaiFre:17}, 
it was set to $0$, and in the remaining 
three algorithms, we took $1/n$.

In Fig.~\ref{fig:single_update}, the error 
$\|x(k)-x^*\|_1$
from the true PageRank vector $x^*$ is plotted in the
logarithmic scale for all five algorithms. 
We must highlight that 
while most of them decrease exponentially fast, 
our proposed method is by far the fastest. 
The plot is cut at the error level of $10^{-2}$, but 
in fact, the decrease in error continues at this rate. 

\subsubsection{Influence of Initial States}
It may appear that the fast convergence of
the proposed Algorithm~\ref{alg:2}
is due to the restricted choice in the initial values.
Since many pages in the web take very small values,
assigning the smallest possible value $m/n$ as the
initial values may be advantageous. 
To check this point, 
we also ran simulations of other methods by assigning 
$m/n$ as the initial states to pages
having especially small PageRanks; other pages received
values by equally dividing the remaining PageRank.
However, the
results did not exhibit major changes in the responses,
at least at the scale visible in plots similar to 
Fig.~\ref{fig:single_update}. Thus, we conclude that
at least for this example, the influence of the
initial states seems very limited. 

\subsubsection{Clustering-Based Method}
\begin{figure}[t]
  \centering
  \subfigure[]{%
      \label{fig:agg_sim1}
       \includegraphics[width=8.3cm,height=5.15cm]{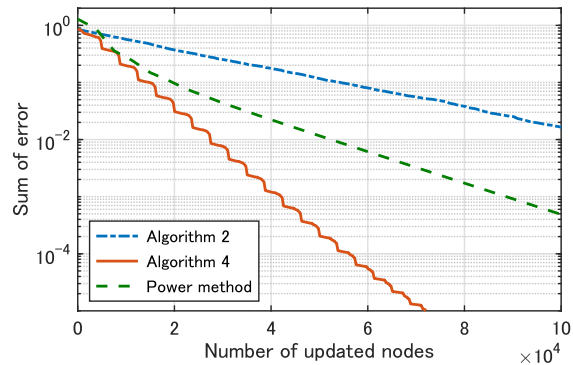}
       \vspace*{-2cm}}
  \subfigure[]{%
           \label{fig:agg_sim2}
           \includegraphics[width=8.3cm,height=5.15cm]{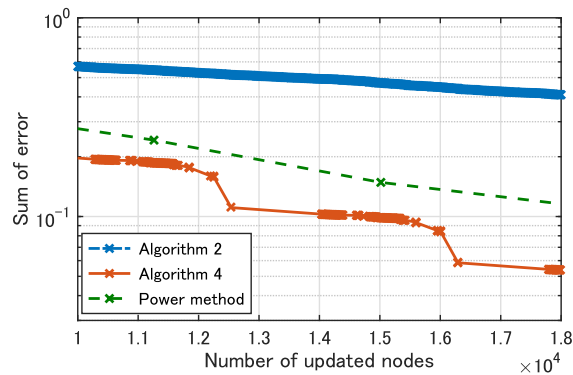}
           \vspace*{-2cm}}
 \caption{(a)~Time responses of errors: 
  Algorithm~\ref{alg:2}, 
  aggregation-based 
  Algorithm~\ref{alg:aggregation}, and 
  the power method. (b)~Enlarged with markers placed at time instants
            when group-based updates are made.}
 \label{fig:agg_sim}
\end{figure}

Finally, in Fig.~\ref{fig:agg_sim1},
we compare the performance of the two proposed
algorithms and the (centralized) power method.
To make the comparison fair,
the horizontal axis is taken as the number of updated nodes.
The page selection for 
Algorithm~\ref{alg:2} is
by randomization and 
that for Algorithm~\ref{alg:aggregation} is
periodic (i.e., selecting groups as $1,2,\ldots,n,1,2\ldots$).
Though the power method
is faster than Algorithm~\ref{alg:2}, 
and also comparable with Algorithm~\ref{alg:aggregation}
in the very beginning, 
the clustering-based method
shortly catches up and
shows faster convergence. 

In Fig.~\ref{fig:agg_sim2}, we show an 
enlarged version of Fig.~\ref{fig:agg_sim1} with markers $\times$ put
at the times when updates by groups took place.
It shows how for Algorithm~\ref{alg:aggregation},
the error decreases
when certain groups make updates in their state values. 
In fact, groups~282 and~290 mentioned above have
major contributions here. This is likely because 
their member pages take large values. 


\section{Conclusion}\label{sec:concl}

In this paper, we have developed a new class of distributed 
algorithms for the computation of PageRank based on 
a reinterpretation of its definition. 
We first have introduced two types of distributed algorithms,
synchronous and asynchronous based on gossipping. 
Their exponential convergence properties have been established, and 
the relation of the proposed algorithms to those
in the literature has been discussed. 
The second part of the paper has been devoted to their
extensions to the case with multiple updates and grouping. 
We have shown that our algorithms 
exhibit superior performance through simulations using
real web data. 
In future research, we will further analyze the 
convergence speeds of the algorithms and employ 
other schemes for page selections.



\appendix
\subsection{Proof of Lemma~\ref{generalized_monot}}
\label{appendix:1}

By \eqref{eqn:alg4:QRS} and \eqref{general_q}, 
we can easily show $z(k)\geq0$ and
$0\leq x(k)\leq x(k+1)$.
Thus, in the remaining, we must show $x(k)\leq x^*$. 
By using the state $x_{\rm{s}}(k)$ of the synchronous 
update scheme \eqref{eqn:synch}, 
it suffices to show 
\begin{equation}
 x(k)\leq x_{\rm{s}}(k)~~\text{for $k\geq 0$}. 
 \label{eqn:unif_thm:2}
\end{equation}
Then, by Lemma~\ref{lemm:synch}\,(ii), 
we obtain $x(k)\leq x_{\rm{s}}(k)\leq x^{*}$.

For $k=0$, we have $x(0)= x_{\rm{s}}(0)$, and 
thus \eqref{eqn:unif_thm:2} holds.
For $k\geq 1$, by \eqref{general_q}, 
the closed-form solution of $z(k)$ is given by
\begin{equation*}
  z(k)
   = Q_{\phi(k-1)} Q_{\phi(k-2)} \cdots Q_{\phi(0)} z(0)
\end{equation*}
and that of $x(k)$ as
\begin{align*}
 &x(k) 
  = z(0) 
     + \displaystyle 
        \sum_{t=0}^{k-1}
          R_{\phi (t)}z(t)\notag\\
  &~~= \frac{m}{n}\mathbf{1}_n 
       + \sum_{t=0}^{k-1}
          R_{\phi (t)}Q_{\phi (t-1)}Q_{\phi (t-2)}
            \cdots Q_{\phi (0)}\frac{m}{n}\mathbf{1}_n .
\end{align*}
For $k\geq 0$, let 
\begin{equation*}
 \widetilde{Q}_{\phi}(k)
   := R_{\phi(k)} Q_{\phi(k-1)} Q_{\phi (k-2)}
           \cdots Q_{\phi(0)}.
\end{equation*}
Then, we have
\begin{equation}
 x(k)
  = \frac{m}{n}\mathbf{1}_n
      + \sum_{t=0}^{k-1}
         \widetilde Q_{\phi}(t)
           \frac{m}{n}\mathbf{1}_n.
\label{general_x}
\end{equation}

Thus, for establishing \eqref{eqn:unif_thm:2},
by \eqref{eqn:x_s} of Lemma~\ref{lem:synch}, we must show 
for each $k>0$
\begin{equation}
  \sum_{t=0}^{k-1}
     \bigl[
        \widetilde Q_{\phi}(t)  
     \bigr]_{ij}
  \leq \sum_{t=0}^{k-1}
         \bigl[
            \widetilde Q_{s}(t)  
         \bigr]_{ij}.
\label{general_q_neq1}
\end{equation}
Since the matrices $\widetilde Q_{\phi}(t)$ and 
$\widetilde Q_{s}(t)$ are nonnegative, 
more specifically, it suffices to show 
that the following inequality holds element-wise:
\begin{equation}
     \bigl[\widetilde Q_{\phi}(t)  \bigr]_{ij}
  \leq 
         \bigl[\widetilde Q_{s}(t)  \bigr]_{ij}
 ~~\text{for $i,j$ and
          $t=0,1,\ldots,k-1$}.
\label{general_q_neq}
\end{equation}
The terms in the right-hand side 
of \eqref{general_q_neq} are written out 
in \eqref{eqn:Qs_ij} of Lemma~\ref{lem:synch}.
Thus, we will obtain the expressions for
$\bigl[ \widetilde{Q}_{\phi}(t) \bigr]_{ij}$
on the left-hand side.
Here, due to the relations in \eqref{eqn:alg4:relation1}
and \eqref{eqn:alg4:relation2}, we have
\begin{align*}
  \widetilde Q_{\phi}(t)	
    &= R_{\phi (t)}Q_{\phi (t-1)}Q_{\phi (t-2)}
        \cdots Q_{\phi (0)}\\
    &= R_{\phi(t)}
         \left( 
            S_{\phi(t-1)}+R_{\phi(t-1)} 
         \right)\cdots
         \left( 
            S_{\phi(0)} + R_{\phi(0)} 
         \right)\\
    &= \bigg( 
         \sum_{i\in\phi(t)}R_i 
       \bigg)
       \bigg( 
         S_{\phi(t-1)}
          + \sum_{i\in\phi(t-1)}R_i 
       \bigg)\\
    &\hspace*{2.5cm}\mbox{}
      \cdots
       \bigg( 
          S_{\phi(0)} + \sum_{i\in\phi(0)}R_i 
       \bigg).
\end{align*}
This indicates that $\widetilde Q_{\phi}(t)$
can be written as the sum of all matrices which 
are products of $t+1$ matrices of the form
\begin{align}
  &R_{m_l}S_{\phi(k_l-1)}
      \cdots S_{\phi(k_{l-1}+1)}\nonumber\\
  &\hspace*{.5cm}\cdot R_{m_{l-1}} S_{\phi(k_{l-1}-1)}
          \cdots S_{\phi(k_{l-2}+1)}\nonumber\\
  &\hspace*{1cm}\cdots
      R_{m_{1}} S_{\phi(k_{1}-1)}
            \cdots S_{\phi(0)},
\label{general_tenkai}
\end{align}
where the indices $k_1,\ldots,k_l$ are taken such that
$k_l=t$ and $0\leq k_1<k_2<\cdots <k_l$; moreover, 
$m_i$ are taken such that $m_i\in \phi(k_i)$,
$i=1,2,\ldots,l$.

From \eqref{eqn:alg4:relation3},
we have that the matrix product 
in \eqref{general_tenkai} is equal to 
either a nonzero matrix
of the form $R_{m_l}R_{m_{l-1}}\cdots R_{m_{1}}\geq 0$ 
or a zero matrix.
It becomes a zero matrix if the chosen 
sequence $\phi$ and the indices $m_{1},\ldots,m_{l}$ 
satisfy at least one of the following conditions:
\begin{itemize}
 \item $m_{1}$ is contained in one of the sets $\phi(k_{1}-1),\ldots,\phi(0)$;
 \item $m_{2}$ is contained in one of the sets $\phi(k_{2}-1),\ldots,\phi(k_1+1)$;
 \item $\ldots$
 \item $m_l$ is contained in one of the sets $\phi(k_l-1),\ldots,\phi(k_{l-1}+1)$.
\end{itemize}
If none of the conditions holds, then 
the matrix product in \eqref{general_tenkai} becomes equal to
$R_{m_{l}}R_{m_{l-1}}\cdots R_{m_{1}}$. 

Next, we reduce the expression of 
$R_{m_{l}}R_{m_{l-1}}\cdots R_{m_{1}}$ 
to a product of scalars and matrices. 
Here, we can use the formula
\begin{align*}
  R_{i} R_{j}
   &= q_{ij} R_{j}^{(i)},
\end{align*}
where $R_{j}^{(i)}$ is nonzero only in the $j$th column as
\begin{equation}
  R_{j}^{(i)}
    := \begin{bmatrix}
            {0}_n & {0}_n & \cdots& {0}_n & q_i
             & {0}_n & \cdots & {0}_n 
         \end{bmatrix}.
 \label{eqn:R_ji}
\end{equation}
We also need another formula that holds for arbitrary $i,j,m$:
\[
  R_{j}^{(i)} R_m = q_{jm} R_{m}^{(i)}.
\]
By repeatedly applying these formulae, we obtain
\begin{align}
 &R_{m_{l}}R_{m_{l-1}}\cdots R_{m_{1}} \notag\\
 &\hspace*{1cm}
   = q_{m_{l}m_{l-1}}q_{m_{l-1}m_{l-2}}\cdots 
     q_{m_{2}m_{1}}
     R_{m_{1}}^{(m_{l})}.
\label{generalized_rr}
\end{align}
Here, the $(i,j)$th element can be expressed as
\begin{align*}
  &\left[ 
      R_{m_{l}}R_{m_{l-1}}\cdots R_{m_{1}} 
   \right]_{ij}\\
  &~~= q_{m_{l}m_{l-1}}q_{m_{l-1}m_{l-2}}\cdots   
       q_{m_{2}m_{1}}\left[R_{m_{1}}^{(m_{l})}\right]_{ij}\\
  &~~= \begin{cases}
       q_{im_{l}}q_{m_{l}m_{l-1}}\cdots q_{m_{2}j}
         & \text{if $m_1= j$},\\
       0 & \text{otherwise}.
     \end{cases}
\end{align*}

So far, we have shown that the summation
$\sum_{t=0}^{k-1}\left[  \widetilde{Q}_{\phi}(t) \right]_{ij}$
in the left-hand side of \eqref{general_q_neq}
can be described as a sum of the terms 
$q_{im_l}q_{m_lm_{l-1}}\cdots q_{m_2j}$ 
using the set of nodes, 
$\left( m_2,\ldots,m_l \right)\in\mathcal{V}^{l-1}$, 
$2\leq l\leq k-1$.

To establish the inequality in \eqref{general_q_neq},
we must prove that for each $i,j,k$ and each node sequence
$\left( m_2,\ldots,m_{\ell} \right)\in\mathcal{V}^{\ell-1}$, 
the number of sequences of time $0\leq k_1<\cdots<k_{\ell}\leq k-1$ satisfying the 
following conditions is at most one:
\begin{equation}
  \left[ 
     R_{\phi (k_{\ell})}R_{\phi (k_{\ell-1})}\cdots R_{\phi(k_1)}   
  \right]_{ij}
   = q_{im_{\ell}}q_{m_{\ell} m_{\ell-1}}\cdots q_{m_2j}.
\label{general_meidai}
\end{equation}
We can find the times $k_1,\ldots,k_{\ell}$ so that the left-hand side of
\eqref{general_meidai} becomes nonzero in a unique manner by the given
sequence of $\phi(k)$ through the following procedure:
\begin{itemize}
 \item $k_1$ is the smallest $t\geq0$ such that $m_1=j\in\phi(t)$;
 \item $k_2$ is the smallest $t>k_1$ such that $m_2\in\phi(t)$;
 \item $\ldots$
 \item $k_{\ell}$ is the smallest $t>k_{\ell-1}$ such that $m_{\ell}\in\phi(k)$. 
\end{itemize}
If there are $k_1,\ldots,k_{\ell}$ such that
$j\in\phi(k_1)$, $m_2\in\phi(k_2)$, $\ldots$, 
$m_{\ell}\in\phi(k_{\ell})$, 
then the corresponding term in 
\eqref{general_meidai} becomes zero. 
In conclusion, there is at most one combination of
times $k_1,\ldots,k_{\ell}$ for which 
\eqref{general_meidai} holds.

To summarize, we have that for each $2\leq \ell\leq k-1$,
the left-hand side of \eqref{general_q_neq}
contains at most one term expressed as
$q_{i m_{\ell}}q_{m_{\ell}m_{\ell-1}}\cdots q_{m_2j}\geq0$.
As shown in \eqref{eqn:Qs_ij}, 
the right-hand side always contains one 
term $q_{im_{\ell}}q_{m_{\ell}m_{\ell-1}}\cdots q_{m_2j}\geq0$.
Therefore, for each $i,j,t$,
the inequality \eqref{general_q_neq} holds. 
This implies $x(k)\leq x^*$, and consequently,
we arrive at $x(k)\leq x(k+1)\leq x^*$.
\hfill\mbox{$\blacksquare$}

\subsection{Proof of Lemma~\ref{aggregation_monot}}
\label{appendix:2}

We can write $z(k)$ as
\begin{align}
 z(k)
  &= S_{\mathcal{V}_{\psi(k-1)}}
       \left(
          I + \widehat{R}_{\mathcal{V}_{\psi}(k-1)}
       \right)
     S_{\mathcal{V}_{\psi(k-2)}}
       \left(
         I+ \widehat{R}_{\mathcal{V}_{\psi(k-2)}} 
       \right)\notag\\
  &\hspace*{2cm}\mbox{}
     \cdots 
       S_{\mathcal{V}_{\psi(0)}}
        \left(
           I+\widehat{R}_{\mathcal{V}_{\psi(0)}}
        \right)z(0). 
\label{aggr_z}
\end{align}
Moreover, $x(k)$ can be written as
\begin{align}
 x(k) 
  &= z(0) 
     + \sum_{t=0}^{k-1}
         \widehat{R}_{\mathcal{V}_{\psi(t)}} z(t)
     \nonumber\\
  &= x(0) 
     + \sum_{t=0}^{k-1}
         \widehat{R}_{\mathcal{V}_{\psi(t)}}
            S_{\mathcal{V}_{\psi(t-1)}}
         \left(
            I+\widehat{R}_{\mathcal{V}_{\psi(t-1)}}
         \right)\notag\\
  &\hspace*{2.2cm}\cdots 
         S_{\mathcal{V}_{\psi(0)}}
         \left(
           I+\widehat{R}_{\mathcal{V}_{\psi(0)}}
         \right)z(0)\notag\\
  &=: \frac{m}{n}\mathbf{1}_n
     + \sum_{t=0}^{k-1}
          \widehat{Q}_{\psi(t)} \frac{m}{n}\mathbf{1}_n.
\label{aggr_x}
\end{align}
Since $Q_{\mathcal{V}_{\psi(t)}}$, 
$R_{\mathcal{V}_{\psi(t)}}$, $S_{\mathcal{V}_{\psi(t)}}$, and
$z(0)$ are all nonnegative, 
$z(k)$ is nonnegative as well. 
Therefore, the term 
$\widehat{R}_{\mathcal{V}_{\psi(k)}}z(t)$
appearing in $x(k)$ is nonnegative. 
Hence, it holds $x(k)\leq x(k+1)$. 

It thus suffices to show that for each $k$
\begin{equation}
  \sum_{t=0}^{k-1}
   \left[
     \widehat{Q}_{\psi(t)} 
   \right]_{ij}
  \leq \sum_{t=0}^{\infty}
    \left[\widetilde Q_{s}(t)  \right]_{ij}
  ~~\text{for $i,j$}.
\label{aggr_proof_1}
\end{equation}
Note that the summation on the right-hand side has
infinite terms, which is different from the relations
used in other proofs such as that of Theorem~\ref{unif_thm}.
We can however still use the expression 
for $\widetilde Q_{s}(t)$
given in \eqref{eqn:Qs_ij} of Lemma~\ref{lem:synch}.

In what follows, we show that 
the left-hand side of \eqref{aggr_proof_1}
is the sum of terms written only in the form of
$q_{im_{\ell}}q_{m_{\ell}m_{\ell-1}}\cdots q_{m_1j}\geq 0$
and, moreover, these terms are all distinct in 
that each term is different from others.
That is, the terms appearing in the left-hand side
form a subset of those in the right-hand side of
\eqref{aggr_proof_1}, confirming the inequality. 

In \eqref{aggr_x}, 
observe that $\widehat{Q}_{\psi(t)}$ consists of 
terms only of the form
\begin{equation}
 R_{\mathcal{V}_{\psi(t-1)}}^{\alpha_t + 1}
  S_{\mathcal{V}_{\psi(t-2)}}
  R_{\mathcal{V}_{\psi(t-2)}}^{\alpha_{t-1}}\cdots  
  S_{\mathcal{V}_{\psi(0)}}
  R_{\mathcal{V}_{\psi(0)}}^{\alpha_0},
\label{aggr_q_elem}
\end{equation}
where $\alpha_{t},\ldots, \alpha_{0}$ are nonnegative integers. 
Here, note that by the definitions of 
$R_{\mathcal{V}_h}$ and $S_{\mathcal{V}_h}$,
this product is either a zero matrix 
or a product of matrices $R_{\mathcal{V}_h}$.
Moreover, the $(i,j)$th element of the matrix in 
\eqref{aggr_q_elem} is nonnegative 
and is a summation of terms only expressed as
$q_{i m_{\ell}}q_{m_{\ell} m_{\ell-1}}\cdots q_{m_1j}$
for $\ell> 0$.

Based on this observation, for establishing
\eqref{aggr_proof_1}, we must show that 
for each term expressed as 
$q_{im_{\ell}}q_{m_{\ell}m_{\ell-1}}\cdots q_{m_1j}$,
there is only one matrix product in the form
of \eqref{aggr_q_elem} whose $(i,j)$th element
contains the term.

For each page $m$, denote by $g(m)$ the index of the
group to which it belongs. 
Then, it can be confirmed that 
a term expressed as
$q_{im_{\ell}}q_{m_{\ell}m_{\ell-1}}\cdots q_{m_1j}$ is
contained in
\begin{align}
 \left[
   R_{\mathcal{V}_{ g\left( m_{\ell} \right) }}
   R_{\mathcal{V}_{ g\left( m_{\ell-1} \right) }}\ldots 
   R_{\mathcal{V}_{ g\left( j \right) }}  
 \right]_{ij}.
\label{aggr_qq}
\end{align}
To find the product of the form \eqref{aggr_q_elem} 
that contains the term
$q_{im_{\ell}}q_{m_{\ell}m_{\ell-1}}\cdots q_{m_1j}$,
we consider inserting matrices 
$R_{\mathcal{V}_h}$ and $S_{\mathcal{V}_h}$ 
between the matrices in \eqref{aggr_qq} anywhere except on the left side of
$R_{\mathcal{V}_{ g\left( m_{\ell} \right) }}$.

First, if we insert any matrix $R_{\mathcal{V}_h}$
in \eqref{aggr_qq}, then the term 
$q_{im_{\ell}}q_{m_{\ell}m_{\ell-1}}\cdots q_{m_1j}$
will not be present any more. 
Next, we consider inserting 
$S_{\mathcal{V}_h}$ in \eqref{aggr_qq}. 
By definition, it holds 
\begin{align}
  R_{\mathcal{V}_{h_1}}S_{\mathcal{V}_{h_2}}	
   &= \begin{cases} 
         0 & \text{if $h_1=h_2$},\\
	 R_{\mathcal{V}_{h_1}} & \text{if $h_1\neq h_2$}.
      \end{cases}
\end{align}
Thus, if $S_{\mathcal{V}_{h_1}}$ is inserted
on the right side of a matrix $R_{\mathcal{V}_{h_2}}$
with $h_1\neq h_2$, then the term 
$q_{im_{\ell}}q_{m_{\ell}m_{\ell-1}}\cdots q_{m_1j}$
will remain in the product.
This fact indicates that 
$S_{\mathcal{V}_{ g\left(m_{u}  \right) }}$ cannot 
be inserted between matrices 
$R_{\mathcal{V}_{g(m_u)}}$ and $R_{\mathcal{V}_{g(m_{u-1})}}$ 
whose indices satisfy $g(m_{u})=g(m_{u-1})$.

Now, we can find the specific matrix product of the
form \eqref{aggr_q_elem} containing the term 
$q_{im_{\ell}}q_{m_{\ell}m_{\ell-1}}\cdots q_{m_1j}$. 
First, the product should contain $\ell+1$ matrices
of the type $R_{\mathcal{V}_h}$. Thus, the product 
would be written as
\begin{align}
  &R_{\mathcal{V}_{ g\left( m_{\ell} \right) }}
   \underbrace{\cdots}_{0 \rm{\;or\; more\;} 
   S_{\psi}}R_{\mathcal{V}_{ g\left( m_{\ell-1} \right) }}
   \underbrace{\cdots}_{0 \rm{\;or\; more\;} 
   S_{\psi}}\cdots
  \nonumber\\
  &R_{\mathcal{V}_{ g\left( m_{1} \right) }}
   \underbrace{\cdots}_{0 \rm{\;or\; more\;} 
   S_{\psi}}R_{\mathcal{V}_{ g\left( j \right) }}
   \underbrace{\cdots}_{0 \rm{\;or\; more\;} 
   S_{\psi}}.
\label{aggr_rs}
\end{align}
For the given sequence $\psi(0),\ldots,\psi(k-1)$, 
this product is a nonzero matrix if there exists 
a sequence $0\leq i_0 \leq \cdots \leq i_{\ell}\leq k-1$ 
of time instants satisfying the following:
\begin{itemize}
 \item $i_0$ is the smallest $k\geq 0$ such that
       $g\left( j \right)\in\psi(k)$;
 \item $i_1$ is equal to $i_0$ if 
       $g\left( m_1 \right)=g\left( j \right)$;
       otherwise, it is the smallest $k>i_0$ such that 
       $g\left( m_1 \right)\in\psi(k)$;
 \item $\cdots$
 \item $i_{\ell}$ is equal to $i_{\ell-1}$ if 
       $g\left( m_{\ell}\right)=g\left( m_{\ell-1} \right)$;
       otherwise, it is the smallest $k>i_{\ell-1}$ such that
       $g\left( m_{\ell} \right)\in\psi(k)$.
\end{itemize}
It is clear that if such a sequence $i_0,\ldots, i_{\ell}$
exists, then it is unique. 

Therefore, we conclude that the left-hand side of
\eqref{aggr_proof_1} consists of terms only in the form
$q_{im_{\ell}}q_{m_{\ell}m_{\ell-1}}\cdots q_{m_1j}$
and they are all distinct. 
\hfill\mbox{$\blacksquare$}

\bibliographystyle{plain}
\bibliography{BibDataBase_jun18.bib}

\begin{thebibliography}{10}

\bibitem{AldKha:91}
R.{\;}W. Aldhaheri and H.{\;}K. Khalil.
\newblock Aggregation of the policy iteration method for nearly completely
  decomposable {M}arkov chains.
\newblock {\em IEEE Trans.\ Auto\-m.\ Contr.}, 36:178--187, 1991.

\bibitem{AvrLitNem:07}
K.~Avrachenkov, N.~Litvak, D.~Nemirovsky, and N.~Osipova.
\newblock Monte {C}arlo methods in {P}age\-{R}ank computation: {W}hen one
  iteration is sufficient.
\newblock {\em SIAM J.\ Numer.\ Anal.}, 45:890--904, 2007.

\bibitem{BiyArc:08}
E.~B{\i}y{\i}k and M.~Arcak.
\newblock Area aggregation and time-scale modeling for sparse nonlinear
  networks.
\newblock {\em Syst.\ \& Cont.\ Lett.}, 57:142--149, 2008.

\bibitem{BriPag:98}
S.~Brin and L.~Page.
\newblock The anatomy of a large-scale hypertextual {Web} search engine.
\newblock {\em Computer Networks \& ISDN Systems}, 30:107--117, 1998.

\bibitem{BroLem_infret:06}
A.~Z. Broder, R.~Lempel, F.~Maghoul, and J.~Pedersen.
\newblock Efficient {P}age{R}ank approximation via graph aggregation.
\newblock {\em Inform.\ Retrieval}, 9:123--138, 2006.

\bibitem{Bullo:18}
F.~Bullo.
\newblock {\em Lectures on Network Systems}.
\newblock CreateSpace, 2018.

\bibitem{ChaHadRab:16}
T.~Charalambous, C.~Hadjicostis, M.~Rabbat, and M.~Johansson.
\newblock Totally asynchronous distributed estimation of eigenvector centrality
  in digraphs with application to the {PageRank} problem.
\newblock In {\em Proc.\ {\rm 55}th IEEE Conf.\ on Decision and Control}, pages
  25--30, 2016.

\bibitem{CsaJunBlo:14}
B.~C. Cs\'{a}ji, R.~M. Jungers, and V.~D. Blondel.
\newblock {PageRank} optimization by edge selection.
\newblock {\em Discrete Applied Mathematics}, 169:73--87, 2014.

\bibitem{DaiFre:17}
L.~Dai and N.~Freris.
\newblock Fully distributed {PageRank} computation with exponential
  convergence.
\newblock {\em arXiv:1705.09927}, 2017.

\bibitem{FerAkiBou:tac12}
O.~Fercoq, M.~Akian, M.~Bouhtou, and S.~Gaubert.
\newblock Ergodic control and polyhedral approaches to {P}age{R}ank
  optimization.
\newblock {\em IEEE Trans.\ Auto\-m.\ Contr.}, 2013.

\bibitem{FraIshRav:15}
P.~Frasca, H.~Ishii, C.~Ravazzi, and R.~Tempo.
\newblock Distributed randomized algorithms for opinion formation, centrality
  computation and power systems estimation: {A} tutorial overview.
\newblock {\em European J.\ Control}, 24:2--13, 2009.

\bibitem{Gleich:15}
D.~F. Gleich.
\newblock {PageRank} beyond the {W}eb.
\newblock {\em SIAM Review}, 57(3):321--363, 2015.

\bibitem{StaCybRes:06}
Statistical Cybermetrics~Research Group.
\newblock {A}cademic {W}eb {L}inkdatabase {P}roject, {U}niv.\ {W}olverhampton,
  {U.K.}
\newblock [Online] Available: http://cybermetrics.wlv.ac.uk/database/, 2006.

\bibitem{HorJoh:85}
R.{\,}A.\ Horn and C.{\,}R.\ Johnson.
\newblock {\em Matrix Analysis}.
\newblock Cambridge Univ.\ Press, 1985.

\bibitem{IshSuz:18}
H.~Ishii and A.~Suzuki.
\newblock Distributed randomized algorithms for pagerank computation: Recent
  advances.
\newblock In T.~Ba\c{s}ar, editor, {\em Uncertainty in Complex Networked
  Systems: In Honor of Roberto Tempo}, pages 419--447. Birkh\"{a}user, 2018.

\bibitem{IshTem:10}
H.~Ishii and R.~Tempo.
\newblock Distributed randomized algorithms for the {P}age{R}ank computation.
\newblock {\em IEEE Trans.\ Auto\-m.\ Contr.}, 55:1987--2002, 2010.

\bibitem{IshTem:14}
H.~Ishii and R.~Tempo.
\newblock The {P}age{R}ank problem, multi-agent consensus and web aggregation:
  {A} systems and control viewpoint.
\newblock {\em IEEE Control Systems Magazine}, 34:34--53, 2014.

\bibitem{IshTemBai:scl12}
H.~Ishii, R.~Tempo, and E.-W. Bai.
\newblock Page{R}ank computation via a distributed randomized approach with
  lossy communication.
\newblock {\em Syst.\ \& Cont.\ Lett.}, 61:1221--1228, 2012.

\bibitem{IshTemBai:tac12}
H.~Ishii, R.~Tempo, and E.-W. Bai.
\newblock A web aggregation approach for distributed randomized {P}age{R}ank
  algorithms.
\newblock {\em IEEE Trans.\ Auto\-m.\ Contr.}, 57:2703--2717, 2012.

\bibitem{LagZacDab:17}
C.~M. Lagoa, L.~Zaccarian, and F.~Dabbene.
\newblock A distributed algorithm with consistency for {P}age{R}ank-like linear
  algebraic systems.
\newblock In {\em Proc.\ {\rm 20}th IFAC World Congress}, pages 5339--5344,
  2017.

\bibitem{LanMey:06}
A.{\;}N. Langville and C.{\;}D. Meyer.
\newblock {\em Google's PageRank and Beyond: The Science of Search Engine
  Rankings}.
\newblock Princeton University Press, 2006.

\bibitem{LeiChe:15}
J.~Lei and H.-F. Chen.
\newblock Distributed randomized {P}age{R}ank algorithm based on stochastic
  approximation.
\newblock {\em IEEE Trans.\ Auto\-m.\ Contr.}, 60:1641--1646, 2015.

\bibitem{MaeIshAlg:17}
J.~M. Maestre, H.~Ishii, and E.~Algaba.
\newblock Node aggregation for enhancing {P}age{R}ank.
\newblock {\em IEEE Access}, 5:19799--19811, 2017.

\bibitem{MesEge:10}
M.~Mesbahi and M.~Egerstedt.
\newblock {\em Graph Theoretic Methods in Multiagent Networks}.
\newblock Princeton University Press, 2010.

\bibitem{MonOliGas:18}
E.~Montijano, G.~Oliva, and A.~Gasparri.
\newblock Distributed estimation of node centrality with application to
  agreement problems in social networks.
\newblock In {\em Proc.\ {\rm 57}th IEEE Conf.\ on Decision and Control}, pages
  5245--5250, 2018.

\bibitem{NazPol:11}
A.{\;}V. Nazin and B.{\;}T. Polyak.
\newblock Randomized algorithm to determine the eigenvector of a stochastic
  matrix with application to the {P}age{R}ank problem.
\newblock {\em Automation and Remote Control}, 72:342--352, 2011.

\bibitem{PolTre:12}
B.~T. Polyak and A.~Tremba.
\newblock Regularization-based solution of the {PageRank} problem for large
  matrices.
\newblock {\em Automation and Remote Control}, 73:1877--1894, 2012.

\bibitem{RavFraTem:15}
C.~Ravazzi, P.~Frasca, R.~Tempo, and H.~Ishii.
\newblock Ergodic randomized algorithms and dynamics over networks.
\newblock {\em IEEE Trans.\ Control of Network Syst.}, 2:78--87, 2015.

\bibitem{SarMolPan:15}
A.~Sarma, A.~Molla, G.~Pandurangan, and E.~Upfal.
\newblock Fast distributed {PageRank} computation.
\newblock {\em Theoretical Computer Science}, 561:113--121, 2015.

\bibitem{ShiYuYan:03}
S.~Shi, J.~Yu, G.~Yang, and D.~Wang.
\newblock Distributed page ranking in structured {P2P} networks.
\newblock In {\em Proc.\ Int.\ Conf.\ Parallel Processing}, pages 179--186,
  2003.

\bibitem{SuzIsh:acc18}
A.~Suzuki and H.~Ishii.
\newblock Distributed randomized algorithms for {P}age{R}ank based on a novel
  interpretation.
\newblock In {\em Proc.\ American Control Conf.}, pages 472--477, 2018.

\bibitem{SuzIsh:cdc19}
A.~Suzuki and H.~Ishii.
\newblock {PageRank} computation via web aggregation in distributed randomized
  algorithms.
\newblock Submitted for conference publication, 2019.

\bibitem{TemCalDab_book}
R.~Tempo, G.~Calafiore, and F.~Dabbene.
\newblock {\em Randomized Algorithms for Analysis and Control of Uncertain
  Systems, with Applications, {\rm Second Edition}}.
\newblock Springer, London, 2013.

\bibitem{WanTan:15}
W.~Wang and C.~Y. Tang.
\newblock Distributed estimation of closeness centrality.
\newblock In {\em Proc.\ {\rm 54}th IEEE Conf.\ on Decision and Control}, pages
  4860--4865, 2015.

\bibitem{YouTemQiu:17}
K.~You, R.~Tempo, and L.~Qiu.
\newblock Distributed algorithms for computation of centrality measures in
  complex networks.
\newblock {\em IEEE Trans.\ Auto\-m.\ Contr.}, 62:2080--2094, 2017.

\bibitem{ZhaCheFan:tac13}
W.~Zhao, H.-F. Chen, and H.~Fang.
\newblock Convergence of distributed randomized {P}age{R}ank algorithms.
\newblock {\em IEEE Trans.\ Auto\-m.\ Contr.}, 50:1177--1181, 2013.

\bibitem{ZhuYeLi:05}
Y.~Zhu, S.~Ye, and X.~Li.
\newblock Distributed {P}age{R}ank computation based on iterative
  aggregation-disaggregation methods.
\newblock In {\em Proc.\ {\rm 14}th ACM Conf.\ on Information and Knowledge
  Management}, pages 578--585, 2005.

\end{thebibliography}

\end{document}